\documentclass[lettersize,journal]{IEEEtran}
\usepackage{amsmath,amsfonts}
\usepackage{algorithmic}
\usepackage{algorithm}
\usepackage{array}
\usepackage[caption=false,font=normalsize,labelfont=sf,textfont=sf]{subfig}
\usepackage{textcomp}
\usepackage{stfloats}
\usepackage{url}
\usepackage{verbatim}
\usepackage{graphicx}
\usepackage{cite}

\usepackage{pifont}
\newcommand{\cmark}{\ding{51}}%
\newcommand{\xmark}{\ding{55}}%
\usepackage{tikz}
\newcolumntype{L}[1]{>{\raggedright\let\newline\\\arraybackslash\hspace{0pt}}m{#1}}
\newcolumntype{C}[1]{>{\centering\let\newline\\\arraybackslash\hspace{0pt}}m{#1}}
\newcolumntype{R}[1]{>{\raggedleft\let\newline\\\arraybackslash\hspace{0pt}}m{#1}}
\DeclareMathOperator*{\argmin}{arg\,min}

\begin{document}

\title{Hardware Trojan Detection using Graph Neural Networks}

\author{
    Rozhin Yasaei, ~\IEEEmembership{Student Member,~IEEE,}
    Luke Chen, ~\IEEEmembership{Student Member,~IEEE,}
    Shih-Yuan Yu, ~\IEEEmembership{Student Member,~IEEE,} 
    Mohammad Abdullah Al Faruque, ~\IEEEmembership{Senior Member,~IEEE}
}

\markboth{IEEE Transactions on Computer Aided Design of Integrated Circuits and Systems}%
{Yasaei \MakeLowercase{\textit{et al.}}: Hardware Trojan Detection using Graph Neural Networks}


\maketitle

\begin{abstract}
The globalization of the Integrated Circuit (IC) supply chain has moved most of the design, fabrication, and testing process from a single trusted entity to various untrusted third party entities around the world. The risk of using untrusted third-Party Intellectual Property (3PIP) is the possibility for adversaries to insert malicious modifications known as Hardware Trojans (HTs). These HTs can compromise the integrity, deteriorate the performance, and deny the functionality of the intended design. Various HT detection methods have been proposed in the literature; however, many fall short due to their reliance on a golden reference circuit, a limited detection scope, the need for manual code review, or the inability to scale with large modern designs. We propose a novel golden reference-free HT detection method for both Register Transfer Level (RTL) and gate-level netlists by leveraging Graph Neural Networks (GNNs) to learn the behavior of the circuit through a Data Flow Graph (DFG) representation of the hardware design. We evaluate our model on a custom dataset by expanding the Trusthub HT benchmarks \cite{trusthub1}. The results demonstrate that our approach detects unknown HTs with 97\% recall (true positive rate) very fast in 21.1ms for RTL and 84\% recall in 13.42s for Gate-Level Netlist.
\end{abstract}

\begin{IEEEkeywords}
Hardware Trojan Detection, Security, Graph Neural Network, Golden Reference-Free, Register Transfer Level, Gate-Level Netlist.
\end{IEEEkeywords}

\section{Introduction}
\IEEEPARstart{T}{he} scale and complexity of modern System-on-Chip (SoC) designs have made it increasingly challenging and expensive for chip manufacturers to design, fabricate, and test every component in-house. The time to market pressure and resource constraints have pushed SoC designers to outsource hardware designs and use Third-Party Electronic Design Automation (3P-EDA) tools and Intellectual Property (IP) cores from various vendors worldwide. Using Third-Party IPs (3PIP) can be cost-effective due to the re-usability of IP cores so that chip manufacturers can reallocate their resources to meet market demands. However, the security and trustworthiness of 3PIPs are not always guaranteed, and reliance on untrusted IPs and EDA tools greatly raises the risks of HT insertion by rogue entities in the IC supply chain.  

HT refers to an intentional and malicious modification of an IC that is usually designed to leak the information, change the functionality, degrade the performance, or deny the service of the chip. Due to the wide applications of ICs in military systems, critical infrastructures, medical devices, etc., the consequences of an undetected HT in a chip can be life-threatening. For example, an actual demonstration of the HT threat occurred in 2007, when a suspected nuclear installation in Syria was bombed by Israeli jets because Syrian radar was disabled by a remote kill switch backdoor in its commercial off-the-shelf microprocessor \cite{syrianRadar}. In 2012, an undocumented hardware backdoor was found in the Actel/Microsemi ProASIC3 chips used in military-grade FPGAs \cite{MilitaryChipBackdoor} that allowed the extraction of secret keys, enabling an adversary to modify the chip's configurations and gain full control of the chip. Furthermore, it is projected that the global semiconductor IP market will reach 7.3 Billion by 2025, with a compounded annual growth rate of 5.5\% from 2020-2025 \cite{MarketAnalysis2020-2025}, and the security concerns about untrusted IPs can significantly damage the market.  

\begin{figure*}[htp!]
\centering
\includegraphics[width=0.98\textwidth]{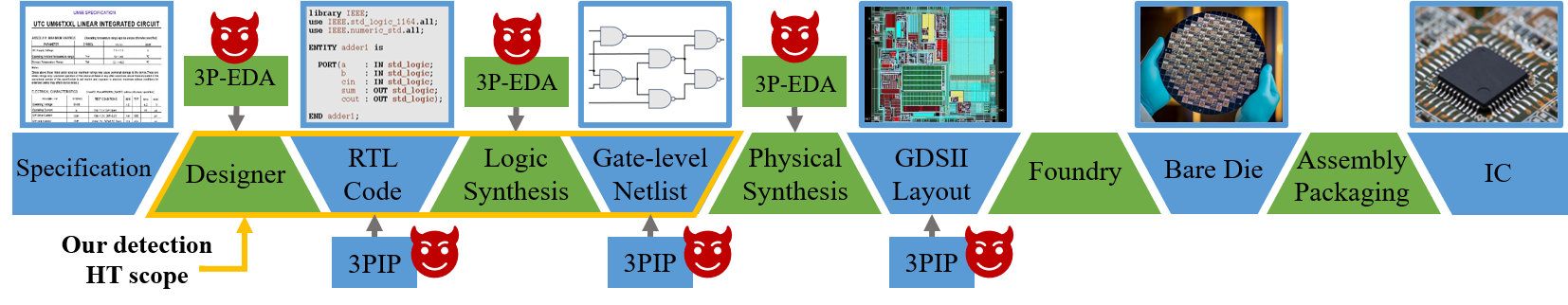}
\caption{Semiconductor supply chain, pre-silicon HT injection points. modify to include netlist as our scope. }
\label{fig:supply-chain}
\vspace{-0.5cm}
\end{figure*}

Figure \ref{fig:supply-chain} shows the typical life cycle of an SoC design, starting from system-level specification to fabrication, in which several stages of the IC supply chain are marked as potential points of HT injection. A rogue in-house designer/team can manually modify or add malicious functions in the hardware design at any design abstraction level (system-level, behavioral-level, and logic level). Moreover, the 3P-EDA tools used for behavioral, logic, and physical synthesis may insert HT into the synthesized RTL and gate-level netlist code. Using untrusted 3PIP in various design stages introduces a means for adversaries to tamper and infect a design with HTs. To ensure the trustworthiness of an SoC design, it is crucial to ascertain the authenticity of 3PIPs. The 3PIPs are typically classified into three categories based on their format; Soft IP (i.e., synthesizable Verilog or VHDL in the RTL format), Firm IP (i.e., gate-level netlist), and Hard IP (i.e., GDSII files and custom physical layout format).

The flexibility of IP cores in higher levels of abstraction makes it easier for the attacker to design and implement various malicious functions. It is crucial to identify HTs early on in the design stages because it becomes increasingly expensive to remove them later. Detecting a few lines of HTs in an industrial-strength IP with thousands and hundreds of thousands of code lines is extremely challenging. Any work requiring manual code review is error-prone, time-consuming, and not scalable. The need for a scalable pre-silicon security verification method is highlighted in many technical documents, such as a recent white paper from Cadence and Tortuga Logic \cite{cadence2019tortuga} because existing solutions fail to safeguard the hardware design from the constantly evolving HTs designed by adversaries. We further elaborate on this necessity through a motivational example.

\subsection{Motivational Example}

The HT detection problem has always been a back-and-forth tug-of-war. HTs are stealthy by design and are composed of a payload and trigger. They are usually very small and inactive with minimal effects on the chip until the trigger circuit is activated under very rare circumstances and triggers the payload to perform its malicious activities. The HT detection problem has always been a back-and-forth tug of war. Whenever new HT detection methods are proposed in the literature, capable of detecting currently known HTs, new HTs are designed to bypass state-of-the-art detection methods. This behavior can be observed by looking at the trend in HT detection over the past decade. One of the earliest defense mechanisms, \cite{UCIxhicks} proposed a novel Unused Circuit Identification (UCI) technique that identifies suspicious circuitry not being used or activated during design verification. However, the authors in \cite{SMC} later designed a new type of HT called Stealthy Malicious Circuits (SMC), which could bypass UCI by hiding HT in nearly-unused logic. Further, FANCI \cite{waksman2013fanci} was successful in detecting SMC by identifying the low control value exhibited by the nearly-unused logic, but it was later defeated by DeTrust \cite{zhang2014detrust}, which designed a new class of HTs with stealthy implicit triggers. Mero\cite{mero} generates test vectors that are capable of activating HTs with low trigger probabilities for detection. Still, it fails to generate test vectors that can activate "hard-to-trigger" HTs with trigger probabilities less than 10\textsuperscript{-6} \cite{saha2015improved}.

Among the more recent works that use Trusthub benchmarks, \cite{rajendran2015detecting} is supposed to detect HTs that leak sensitive data such as secret keys in cryptographic cores as in the AES-T600 benchmark. However, it fails to detect another similar HT benchmark, AES-T700. Later, \cite{rajendran2016formal} identifies different data leaking HTs (e.g., AES-T600 and AES-T700) by adding data leaking as an additional security property for model checking. However, it fails for HTs that execute other malicious functions such as chip degradation (e.g., the AES-500 benchmark) rather than data leakage. A recent paper \cite{fyrbiak2020graph} showed that modeling the hardware design as a graph can be beneficial in the hardware security domain. However, the HT detection scope of \cite{piccolboni2017efficient, fyrbiak2020graph} based on graph similarity algorithms are limited to known HTs in the method's library. There have also been methods that convert the hardware design into a graph representation of the circuit and extract HT features to build a database of HT features for detection \cite{MultilevelFASTrust2018}, but the detection scope is also limited to the HTs in its database.

Despite the various HT detection methods proposed in the literature, the HT problem remains significant as there exists no single method that can detect all different types of HTs. While most detection algorithms can only detect known HTs, new HTs are designed to circumvent existing detection methods. Thus, a new flexible technique for detecting unknown HTs is needed, one that can be expanded as new HTs emerge.
\subsection{Research Challenges}

The existing pre-silicon HT detection methods have several shortcomings: 

1) \textbf{Reliance on golden-reference}: Most HT detection methods rely on a golden HT-free reference circuit to compare the design under test with the golden reference and flag it as HT-infected in the case of deviation. Nevertheless, a golden reference is hard to obtain in practice, especially infeasible at the IP level.

2) \textbf{Limited detection scope}: Some HT detection algorithms are constructed based on a library of known HTs. Consequently, they fall short in detecting unknown HTs. Another set of methods assumes particular properties in the Trojan design of the trigger or payload, limiting the detection scope to specific HTs.

3) \textbf{Manual code review}: Many HT countermeasures mark the parts of the design under test which are prone to HT insertion. However, this does not guarantee that an HT is present/detected, and it would burden the circuit designer with a manual review of the suspicious parts, which is tedious and error-prone for large designs.

4) \textbf{Scalability and complexity issue}: Due to the growing complexity of modern ICs, scalability is an essential feature for any hardware design tool but many current techniques and algorithms are so complicated that they face time or memory issues for large designs.
\subsection{Contributions}

To address these challenges, \textbf{we propose a golden reference-free pre-silicon HT detection approach that takes advantage of state-of-the-art machine learning techniques to learn the circuit behavior and detect the anomalous and malicious presence of Trojan inside the design.} Since hardware is a non-euclidean, structural type of data in nature, we use the graph data structure to represent the hardware design and generate the Data Flow Graph (DFG) for both RTL codes and gate-level netlists. We leverage Graph Neural Networks (GNN) to model the behavior of the circuit. The scalability of our method stems from the fully automated process of graph generation, feature extraction, and detection without any manual workload. Automatic feature extraction is crucial when new HTs are discovered, as our model will be readily updated without the need for defining new properties or introducing additional feature engineering as in previous works. Our contributions are outlined below:
\begin{itemize}
    \item We propose a novel approach for modeling the hardware design to ensure its security. Our methodology models the circuit as its intrinsic representation, a graph, and we leverage GNN to extract the critical features of hardware design and learn its behavior.

    \item We construct two fully automated models for HT detection in the pre-silicon stage. Each model includes a DFG generation pipeline followed by a GNN-based graph learning flow, and it is developed and customized for its target hardware design, either RTL code or gate-level netlist. The models discover even unknown HTs without relying on golden HT-free references or manual code reviews from the circuit designer. Our models are faster than existing methods and scalable for large designs.
    
    \item We create a Trojan DFG dataset consisting of RTL codes and gate-level netlists. We expand the HT-infested RTL benchmarks from Trusthub and perform logic synthesis and optimization to acquire the corresponding gate-level netlist.
    
    \item We survey the pre-silicon HT detection techniques in the literature, analyze state-of-the-art, and make a comprehensive comparison with our approach.
\end{itemize}
\subsection{Threat Model}

This paper aims to determine whether an RTL code or gate-level netlist is infected with a malicious Trojan circuit or not. There is no assumption on the type of HT and the design of trigger or payload. Therefore, our method can detect trigger-based or always active Trojans with a payload circuit to modify functionality, degrade performance, leak confidential data, or deny the service. The only premise is that the HT is inserted in the design stage through the following attack scenarios:
1) A rogue in-house designer intentionally manipulates the RTL/netlist design manually.
2) The 3P-EDA tool used for design/logic synthesis and analysis inserts an HT into the synthesized RTL/netlist code automatically.
3) 3PIP vendor is not trustworthy, and a malicious circuit is hidden in the IP.

\section{Methodology}

Our methodology is built on the premise of the existence of a feed-forward function $f$ that determines whether the hardware design $p$ is infected with a malicious Trojan circuit or not. To approximate the function $f$, our methodology comprises three main steps:

i) Firstly, we extract the DFG $G$ from hardware design $p$ through an automated DFG generation pipeline.

ii) Secondly, the DFGs are passed to a GNN framework that includes graph convolution layers and an attention-based graph pooling layer for graph learning and feature extraction. Further, the graph readout operation generates a vectorized graph embedding, denoted as $\mathbf{h_G}$ based on the discerning features learned by the GNN model.

iii) Lastly, Multi-Layer Perceptron (MLP) is used to perform classification on the graph embedding and output the HT label $y$.The HT label $y$ is the output of function $f$, as given in Equation~\ref{formular1}.
\begin{equation}
\label{formular1}
    y = f(p) = \left \{
    \begin{array}{ll}
         (1,0), &  \text{if the design is HT-infected}\\
         (0,1), &  \text{if the design is HT-free},
    \end{array}
    \right.
\end{equation} 
We describe some background information about GNN and provide more details regarding our GNN model and MLP classifier in Section~\ref{subsec:gnn}. In our approach, we target the hardware design in the pre-silicon stage, including the RTL and gate-level netlist. Although RTL codes and netlists are both represented using a Hardware Description Language (HDL) such as Verilog, they are very different in terms of structure, level of abstraction, and code size (number of signals and operations). Thus, we construct two distinct HT detection models with different graph generation pipelines, as elaborated in Section~\ref{subsec:RTL} and Section~\ref{subsec:netlist}.

\subsection{Graph Convolutional Networks} \label{subsec:gnn}

Many data structures such as social network data or hardware designs can be naturally formulated as graphs. Graph learning encompasses fundamental challenges since the graph size and topology vary among data samples. To address this issue, GNN is introduced. It is a graph-based deep learning model that extracts features from non-Euclidean data structured as a graph. The architecture we use in this paper is inspired by the Spatial Graph Convolution Neural Networks (SGCN). As depicted in Figures~\ref{fig:rtl_model} and \ref{fig:netlist_model}, the architecture of our GNN models for both RTL and netlist HT detection mainly includes convolutional layers, an attention-based pooling layer, a readout unit, and an MLP classifier.

\subsubsection{\textbf{Graph Convolution}}

The input to GNN is a graph $G = (V, E)$ where $E$ is the set of directed edges and $V$ is the set of vertices. The edges are represented as the adjacency matrix $A$ and each node embodies a feature vector $a_v$ that specifies the node attributes. In general, the convolution operation in an SGCN is defined by a node's spatial relations. The spatial convolution has a \textit{message propagation phase} and a \textit{read-out phase. The intuition behind message passing is that nodes pass feature vectors to their immediate graph neighbors and through an iterative process, the information is accumulated as node embeddings. Each iteration is basically one layer of graph convolution and by increasing SGCN layers, nodes can reach further nodes and gather information deeper in the graph. The \textit{message propagation} phase involves two sub-functions: \textbf{AGGREGATE} and \textbf{COMBINE} functions, given by:
\begin{equation}
    a_v^{(l)} = \textbf{AGGREGATE}^{(l)}(\{h_u^{(l-1)}: u \in N(v)\}),
\end{equation}
\begin{equation}
    h_v^{(l)} = \textbf{COMBINE}^{(l)}(h_v^{(l-1)}, a_v^{(l)} ),
\end{equation}
where $N(v)$ is the set of nodes connected to node $v$,  \textbf{AGGREGATE} collects the feature vectors of neighboring nodes to produce an aggregated feature vector $a_v^{(l)}$ for layer $l$. \textbf{COMBINE} will combine the previous node feature $h_v^{(l-1)}$ with $a_v^{(l)}$ to produce the next feature vector $h_v^{(l)}$. The message propagation is carried out for a pre-determined number of $l$ iterations. In our model, the \textbf{AGGREGATE} and \textbf{COMBINE} functions of massage passing are performed by the Graph Convolution Network (GCN)~\cite{kipf2016semi}. Concatenation of aggregated message vectors $a_v^{(l)}$ for all $v \in V$ forms the matrix $\mathbf{X^{(l)}}$, which we call the node embedding matrix. }The GCN layers update the node embeddings for each iteration $l$ of \textit{message propagation} as follows:
\begin{equation}
    \mathbf{X^{(l+1)}} = \sigma(\widehat{D}^{-\frac{1}{2}} \widehat{A} \widehat{D}^{-\frac{1}{2}} \mathbf{X^{(l)}} W^{(l)})
\end{equation}
where $W^l$ is a trainable weight used in the GCN layer. $\widehat{A} = A + I$ is the adjacency matrix of $G$ used for aggregating the feature vectors of the neighboring nodes and $I$ is an identity matrix that adds the self-loop connection to make sure the previously calculated features will also be considered in the current iteration. This self-loop acts similarly to the COMBINE function where the accumulated messages are combined with the previous node feature vector. $\widehat{D}$ is the diagonal degree matrix used for normalizing $\widehat{A}$. $\sigma(.)$ is the activation function such as the Rectified Linear Unit (ReLU). We initialize the embedding  $\mathbf{X}^{(0)}_{i}$ for each node $i \in V$ based on our initial intuition about the graph data in a specific application.
We denote the final propagation node embedding $\mathbf{X}^{(l)}$ as $\mathbf{X}^{prop}$.
Regarding the complexity of our model, GCN complexity grows almost linearly with the number of nodes for sparse graphs, and the circuit connections are sparse,  producing a sparse DFG \cite{liu2020efficient}. The other factor that affects the computation and memory usage is the length of node feature vectors which is a relatively small number (e.i. compared to raw image) in our case.
\subsubsection{\textbf{Attention-based Pooling}}

An attention-based graph pooling layer is then applied to the node embedding $\mathbf{X}^{prop}$. According to ~\cite{knyazev2019understanding}, such an attention-based pooling layer allows the model to concentrate on a local part of the graph and is regarded as part of a unified computational block of a GNN pipeline.
We perform \textit{top-k filtering} on nodes based on the scores predicted from a separate trainable GNN layer~\cite{lee2019self}, as follows:
\begin{equation}
\mathbf{\alpha} = \textbf{SCORE}(\mathbf{X}^{prop}, \mathbf{A}), \text{ }\mathbf{P} = \mathrm{top}_k(\mathbf{\alpha}) 
\end{equation}
where $\mathbf{\alpha}$ denotes the coefficients predicted for nodes by the graph pooling layer. $\mathbf{P}$ represents the indices of the pooled nodes, which are chosen from the top $k$ nodes ranked by $\alpha$. The number $k$ used in \textit{top-k filtering} is computed as $k = pr \times |V|$, where $pr$ is a pre-defined constant called pooling ratio. We consider only a constant fraction $pr$ of the embeddings of the nodes in the DFG to be relevant (i.e., 0.6). The pooling ratio like other hyper-parameters of the GNN model is tuned based on the application through design exploration by the model developer. The node embeddings and edge adjacency information after pooling are denoted by $\mathbf{X}^{pool}$ and $\mathbf{A}^{pool}$ which are calculated as follows:
\begin{equation}
\mathbf{X}^{pool}= (\mathbf{X}^{prop} \odot\mathrm{tanh}(\mathbf{\alpha}))_{\mathbf{P}}, \text{ } \mathbf{A}^{pool} = {\mathbf{A}^{prop}}_{(\mathbf{P},\mathbf{P})} \\
\end{equation}
where $\odot$ represents an element-wise multiplication, $()_{\mathbf{P}}$ refers to the operation that extracts a subset of nodes based on $P$ and $()_{(\mathbf{P},\mathbf{P})}$ refers to the adjacency matrix between the nodes in this subset.

\subsubsection{\textbf{Graph Embedding Generation}}

In the read-out phase, our model aggregates the node embeddings acquired from the graph pooling layer, $\mathbf{X}^{pool}$ and extracts the graph-level features which is a vector called graph embedding $\mathbf{h}_{G}$ for DFG $G$.
\begin{equation}
    \mathbf{h}_{G} = \textbf{READOUT}(\mathbf{X}^{pool})
\end{equation}
Where the \textbf{READOUT} operation may be summation, average, or the maximum of each feature dimension, over all node embeddings, denoted as \textit{sum-pooling}, \textit{mean-pooling}, or \textit{max-pooling} respectively.

\subsubsection{\textbf{Multi-Layer Perceptron Classifier}}

Our GNN model generates a graph embedding for each DFG that represents the essential features of the hardware design. We further process the embedding vector $\mathbf{h}_{G}$ with an MLP layer and a Softmax activation function to produce the final prediction as follows,
\begin{equation}
    \hat{Y} = \mathbf{Softmax}(\mathbf{MLP}(\mathbf{h}_G))
\end{equation}
This layer reduces the number of hidden units used in $\mathbf{h}_{G}$ and produces a two-dimensional output representing the probabilities of both classes (HT-infested or HT-free). Finally, the predicted values in $\hat{Y}$ are normalized using the Softmax function, and the class with the higher predicted value is chosen as the detection result.

To train the model, we compute the cross-entropy loss function, denoted as $H$, between the ground-truth label $Y$ and the predicted label $\hat{Y}$, described as follows:
\begin{equation}
    \argmin H(Y, \hat{Y}) = \argmin \sum_{y_i \in Y, \hat{y_i} \in \hat{Y}} y_i log_e(\hat{y_i})
\end{equation}
The model is trained through an iterative process using the gradient descent method to minimize our cross-entropy loss function.

\subsection{GNN for HT Detection in RTL} \label{subsec:RTL}

\begin{figure*}[t]
\centering
\includegraphics[width=0.98\textwidth]{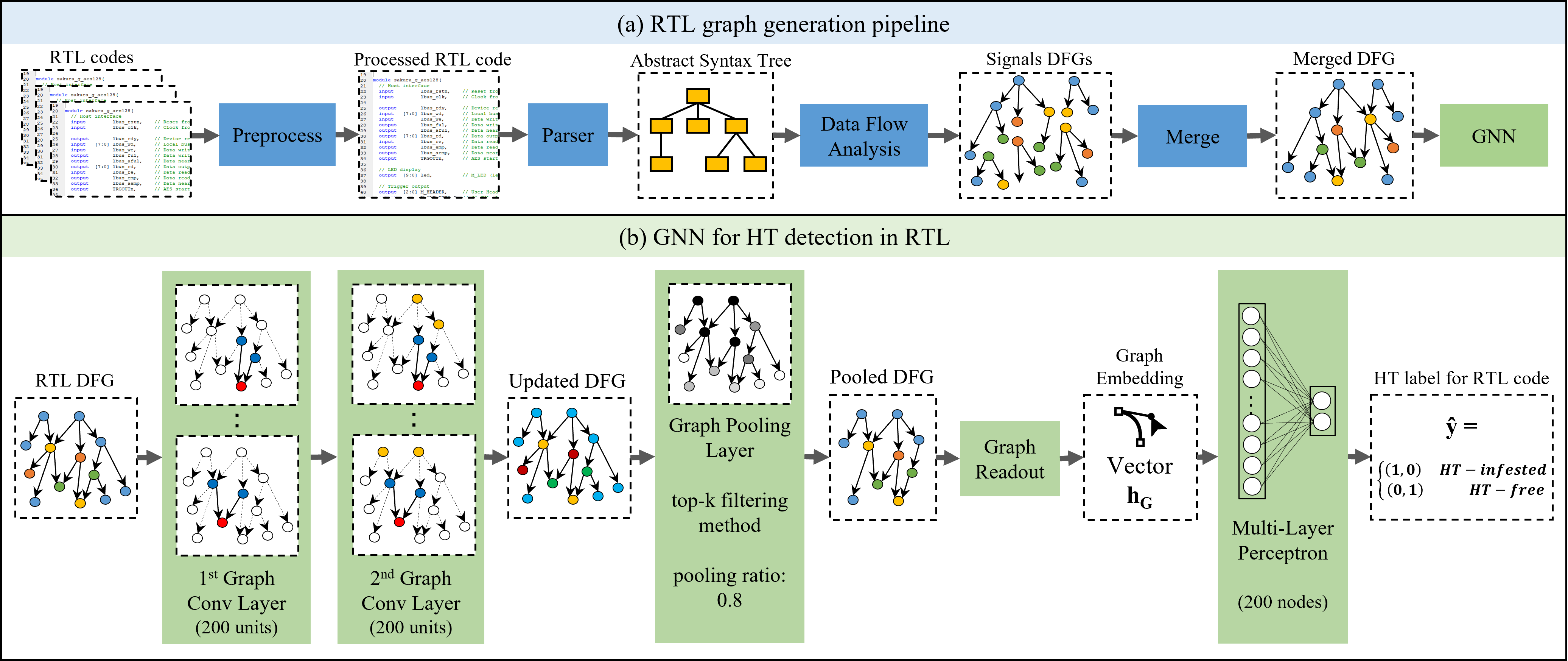}
\caption{The RTL DFG generation pipeline and GNN model for Trojan detection in RTL code.}
\label{fig:rtl_model}
\vspace{-0.5cm}
\end{figure*}

In this section, we explain our first model that processes the hardware design in RTL and determines the presence of HT in the code. The overview of the model is illustrated in Figure~\ref{fig:rtl_model}. The first part of the figure indicates the pipeline to extract data flows of RTL code and structure them as a directional graph, DFG. The next part presents the architecture of the machine learning model that we developed to classify the RTL graphs.

\begin{figure}[t]
\centering
\includegraphics[width=0.46\textwidth]{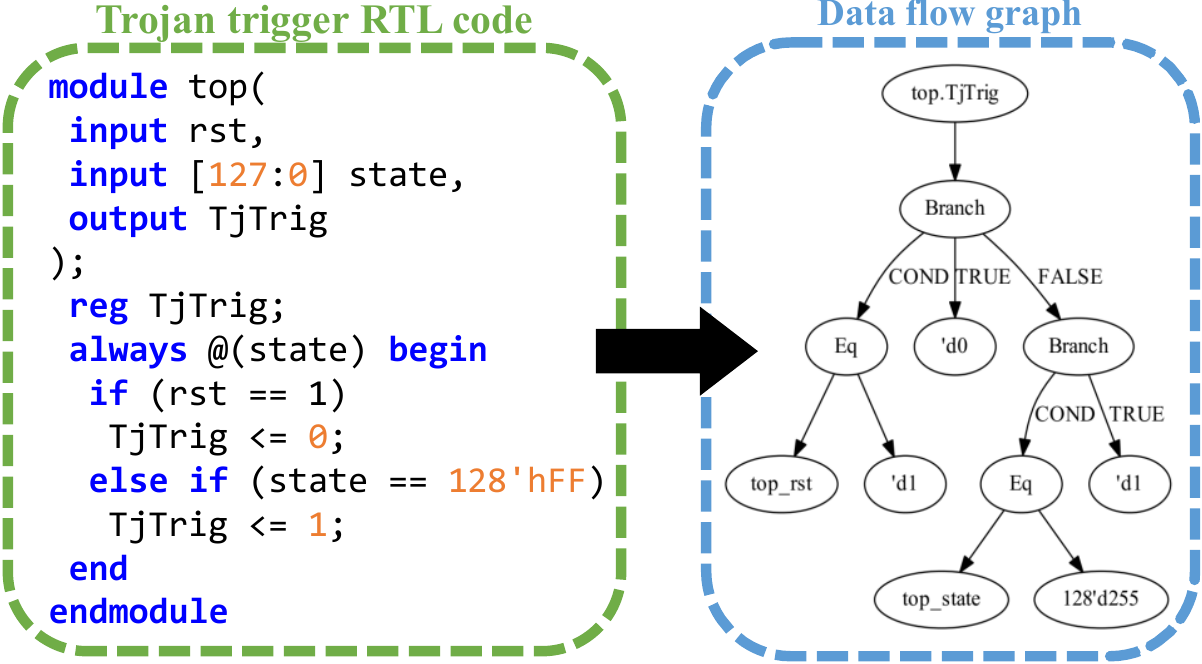}
\caption{The RTL code of a Trojan trigger and its DFG.}
\label{fig:graph}
\vspace{-0.5cm}
\end{figure}
\subsubsection{\textbf{RTL Graph Generation}} 

A graph is a non-euclidean data structure that retains the topological information as well as signals. Similarly, the hardware design describes the circuits in terms of circuit elements and their connections. Given that these are naturally aligned, we convey the hardware design by employing DFG. DFG captures the data dependencies between signals and operations in the circuit and provides a fundamental expression of the computational structure. We develop an automated conversion process of the circuit to DFG through our DFG generation pipeline, as illustrated in Figure~\ref{fig:rtl_model}(a). The automated pipeline comprises several phases: preprocess, parser, data flow analyzer, and merge.

Due to the complexity and size of circuits, a digital circuit is often designed in a hierarchy, with multiple modules in different files. Consequently, the graph generation procedure starts with combining the files and flattening the design to a single RTL code. The preprocessing phase also resolves any syntax incompatibilities (i.e., invalid signal names). Next, we use a hardware design toolkit called Pyverilog \cite{takamaeda2015pyverilog} to parse and analyze the Verilog code. The Pyverilog parser extracts the abstract syntax tree from code and passes it to the data flow analyzer to generate a DFG for each signal in the circuit such that the signal is the root node. To have a single graph representation for the whole circuit, we fuse all the signal DFGs and trim the disconnected sub-graphs and redundant nodes in the merge phase. Figure \ref{fig:graph} exemplifies an RTL code and its corresponding DFG.

In addition to graph topology, node types also contain valuable information for graph learning and are used to create the node feature vectors. We initialize the node feature vector to be the one-hot encoding of the node type. The merged DFG generally contains three categories of node types; operation, constant, and signal. Twenty-eight different types of operation nodes are recognized in RTL graphs, and nodes are tagged by their operation name accordingly (e.i. AND, XOR, etc.). The constant tag indicates that the node represents a number and is tagged as constant regardless of its value. Notably, more variation is observed in signal nodes because their tags are extracted from signal names in the RTL code designated by the designer. We assign signal nodes distinctive tags based on their name in the RTL code because semantic data is concealed in the signal names. After scanning various RTL codes, we create a list of signal names that provide semantic information and are prevalent in different designs. For example, a node name that spells "clk" or "clock" generally refers to the circuit clock. We tag the signals node based on the list, and the nodes with a new signal name not in the list are tagged the same as the "general" signal node. Therefore, the feature vector length is independent of the circuit , and eventually, the length of node feature vectors for RTL graphs is 300 nodes. 
In contrast, such high-level information does not exist in a netlist, and signal names are arbitrary, only to show wires. Thus, we ignore the signal names in the netlist and tag the signals either as input, output, or intermediate in the netlist (elaborated in Section \ref{sec:netlist-norm}).

\subsubsection{\textbf{HT Detection Model for RTL}} 

Figure \ref{fig:rtl_model}(b) demonstrates the architecture of our RTL Trojan detection model based on graph convolutional networks. The mathematical background of GNN and details of each layer are elaborated in Section~\ref{subsec:gnn}, and this section presents the architecture of our customized GNN model for HT detection in RTL. When the DFG of RTL design is generated by the graph generation pipeline, it is passed to the GNN model for graph learning. To be precise, the model inputs are the adjacency matrix of DFG, which indicates the connections among nodes, and a feature vector for each node. Our model architecture includes two convolutional layers with 200 hidden units that perform message passing and modify the node feature vectors. Then, we use a graph pooling layer as our attention mechanism, which pushes the model to focus on the critical nodes. Our pooling method is k-filtering with a ratio of 0.8, which scores the nodes and keeps the nodes with the top 80\% scores. Further, the readout unit extracts the graph embedding vector from the node feature vectors and passes it to the MLP with 200 nodes to classify the RTL code as HT-free or HT-infected. This architecture is achieved after some design space exploration which is discussed in Section~\ref{sec:hyperparameters}.

\begin{figure*}[t]
\centering
\includegraphics[width=0.98\textwidth]{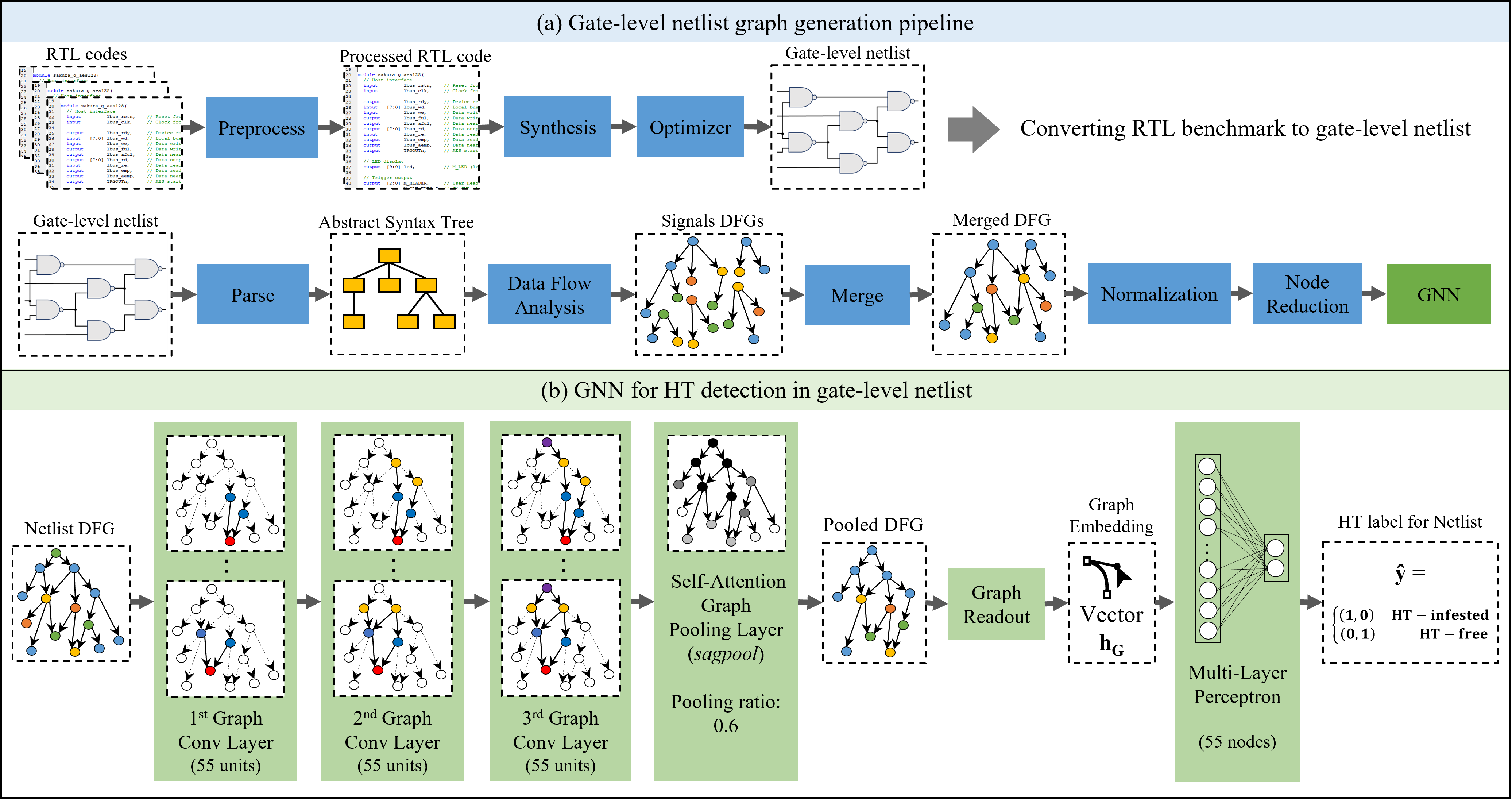}
\vspace{-0.3cm}
\caption{The netlist DFG generation pipeline and GNN model for Trojan detection in gate-level netlist.}
\label{fig:netlist_model}
\vspace{-0.5cm}
\end{figure*}
\subsection{GNN for HT Detection in Netlist}  \label{subsec:netlist}

Gate-level netlist is different from the RTL design as it is closer to the actual hardware implementation and very detailed. The high-level behavior of RTL code vanishes in the netlist, and it is substituted by numerous gates and signals defined to imitate the circuits components and their connection. As a result, the graph representation of netlist is different from RTL with a considerably larger size. Consequently, our RTL HT detection model fails to perform with the expected accuracy for netlist, and we develop a customized graph generation pipeline and GNN model for finding HT in a netlist, as shown in Figure~\ref{fig:netlist_model}.

\subsubsection{\textbf{HT Benchmarks Synthesis to Netlist}}

Following the netlist graph generation procedure in Figure \ref{fig:netlist_model}(a), we employ the open-source RTL synthesis tool, called Yosys \cite{Yosys} to synthesize the RTL HT benchmarks and generate the gate-level netlists. A custom script is used to automate the RTL to gate-level netlist synthesis and perform some additional processing steps to fix the syntax incompatibilities for Pyverilog. The Yosys logic synthesis steps are as follows: 1) The Yosys RTL frontend converts the RTL code into the RTL intermediate Language (RTLIL) internal cell representation. 2) The built-in optimizer is called to remove unused signals and cells, optimize finite state machines, and translate memories to primary memory cells. 3) The built-in technology mapper and ABC tool are used to map the internal cell representation to a custom standard cell library made up of generic gates (AND, OR, XOR, NOT, etc.) and output as a generic gate-level netlist code in Verilog.
One issue during logic synthesis is that the optimization step removes unused signals and cells. According to Yosys, unused signals and cells are defined as those that do not modify an output signal. This becomes a problem as there are HTs that do not alter any output of the circuit. For example, a Trojan is designed to degrade the performance of the chip by causing additional power consumption through excessive shifting. To resolve this issue, we specify in Yosys to not remove any user-defined signals and cells so that it will retain the HT code.

\subsubsection{\textbf{Netlist Graph Generation}} \label{subsec:netlistDFG}

After RTL synthesis, the netlist in Verilog format is generated. The synthesized netlist code also introduces syntax incompatibilities with Pyverilog, so we perform some processing to remove these issues. Finally, the cleaned-up gate-level netlist is passed to the Pyverilog parser. The remaining steps for parsing the AST, generating a DFG for each signal, and merging and trimming are similar to the RTL code analysis to produce a single DFG that represents the whole circuit.
The final DFG for gate-level netlist is a rooted directed graph that shows data dependency from the output signals (the root nodes) to the input signals (the leaf nodes). It is defined as graph $G = (V, E)$ where $E$ is the set of directed edges and $V$ is the set of vertices. We define $V =\{v_1, v_2,..., v_n\}$ where $v_i$ represents signals, constant values, and operations such as xor, and, branch, or branch condition. We define $E ={e_{ij}}$ for all $i, j$ such that $e_{ij} \in E$ if the value of $v_i$ depends on the value of $v_j$, or if the operation $v_j$ is applied on $v_i$.

\subsubsection{\textbf{Netlist Graph Optimization }}

Behavioral characterization of RTL is lost in the netlist due to a lower level of abstraction, and the circuit is described as a detailed, complex sea of gates. Consequently, the netlist graph is significantly larger than its RTL counterpart, thwarting HT detection. For instance, the RTL DFG of AES benchmark, the most complicated benchmark in our dataset, holds on average 13K nodes per graph. Conversely, the AES benchmark for gate-level netlist graph contains 300K nodes per graph on average. The increased graph size drastically expands the memory footprint for both the dataset and the model, leading to memory shortage. Therefore, we reconstruct our dataset and model to reduce the memory footprint.

After reviewing various netlists and their DFGs, we identified unnecessary details in the signals and operations that are ineffective in HT detection. To reduce the graph size, we conduct an automated graph optimization process to eliminate the excessive details of the netlist. Our graph optimizer trims the graph generated by Pyverilog and removes redundant nodes, whose removal does not affect the overall hardware design data-flow representation. For example, the nodes tagged as \textit{concatenation} and \textit{part selection} by Pyverilog provide detailed information that in HT detection application are not required. The \textit{concatenation} node appears for certain node types, such as the logic gates used in the netlist. For example, an output node Y depends on the result of an OR gate operation node between two input signal nodes, A and B. The \textit{concatenation} node exists between the OR gate and the two input signal nodes to signify that the OR gate depends on two signals. When the redundant \textit{concatenation} node is removed, the signal nodes are directly connected to the OR gate without the intermediate \textit{concatenation} node, indicating data dependency without any information loss. 

\begin{figure}[]
\centering
\includegraphics[width=0.45\textwidth]{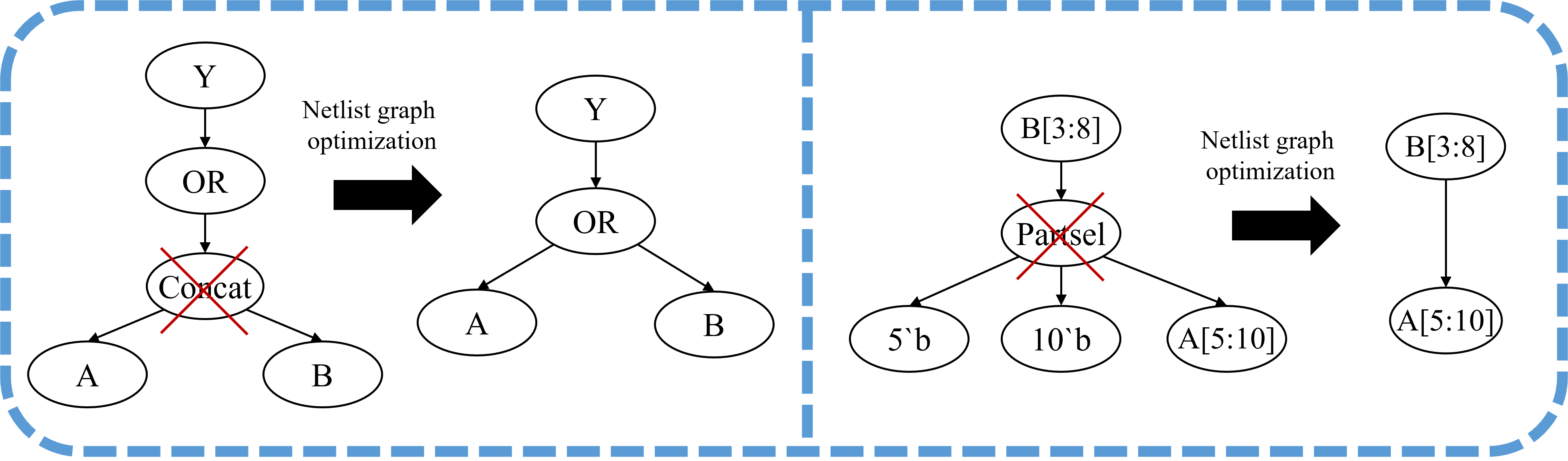}
\caption{Netlist graph optimization by removing \textit{concatenation} and \textit{part selection} nodes.}
\label{fig:opt}
\vspace{-0.5cm}
\end{figure}

The \textit{part selection} node appears when there is a bit-wise assignment. For instance, assigning bits five to ten for a wire node A to bits three to eight for a wire node B. Node A will be connected to node B with a \textit{part selection} node making up of 3 components; the least significant bit of A, the most significant bit of A, and the data signal A itself. We are only concerned about the data dependency in the HT detection application and not the bit-wise detail of the data. Therefore, we exclude the \textit{part selection} nodes and only keep the data signal in the dependency relation. With these changes, we reduced the netlist graph of the AES benchmarks from approximately 300K nodes to 100K nodes. Overall, the average number of nodes for the entire dataset was reduced by 50\%. Figure~\ref{fig:opt} demonstrates examples of netlist DFG optimization through elimination of \textit{concatenation} and \textit{part selection} nodes.
\subsubsection{\textbf{Netlist Graph Normalization} }   \label{sec:netlist-norm}

Although we have significantly reduced the overall size of our netlist dataset, it is still comparably larger than its RTL counterpart. To reduce the memory overhead further, we modify an integral part of the GNN model itself, the node feature vector dimension. The feature vector is used as part of the graph convolution process for node embedding. It is initialized based on intuition about the application and data concealed in the graph nodes. The node feature matrix is an N by D matrix where N is the number of nodes and D is the dimension of the feature vector. We have reduced N in the graph optimization step through node reduction. Therefore, we apply a node normalization step to reduce the feature vector dimension to resolve this issue.

In our application, nodes in a netlist DFG have different types such as signals, constant values, and logic operations. Each signal node in the DFG has a unique name, and tagging the nodes based on their name leads to large feature vectors while the signal names hardly convey any semantic data or behavioral description (i.e. Wire1, Wire2). We normalize the DFGs by generalizing the node tags. Instead of having each unique signal name, constant value, and type of operation as a one-hot encoding, we categorize them into 17 classes.

For example, all nodes with no in-degrees will be assigned as input signals, and all nodes with no out-degree will be designated as output signals. This also applies to the generic gate nodes where each gate is assigned a specific value. Normalization drastically decreases the number of node tags from 300 to only 17, reducing the feature vector dimension and subsequently, the required memory for graph learning.
Moreover, the accurate initialization of the feature vector can improve the performance of GNN to converge fast to high accuracy. Our intuition is that the feature vectors correlate to their node type, and by normalization, we extract this information that is carried over to GNN.

\subsubsection{\textbf{HT Detection Model for Netlist} }

After the netlist DFG is generated and optimized, it is passed to the GNN model along with its normalized node feature vectors. The architecture of our netlist-customized HT detection model is demonstrated in Figure~\ref{fig:netlist_model}(b). We reach this model based on our intuition about the distinct characteristics of netlist after conducting various experiments, some of which are presented in Section~\ref{sec:hyperparameters}. 
The most challenging characteristic of the netlist is the graph size which is relatively larger than the RTL graph for the same circuit, even after optimization. Additionally, it is too detailed and missing the high-level behavioral information of circuits, making the feature extraction challenging. Due to these traits, our netlist model has one more graph convolutional layer than the RTL model (3 layers in total) to perform more in-depth massage passing and feature extraction. Moreover, the number of hidden units is decreased to 55 to minimize resource utilization of the model. Next, we utilize self-attention graph pooling to revise the nodes and focus on those with strong influence. We got the best performance for the pooling ratio of 0.6, which is lower than the RTL ratio. It matches the initial intuition that there are many unnecessary nodes in the netlist graphs which can be omitted through pooling. The rest of the model is very similar to the RTL-customized model, and it creates the graph embedding with a readout unit followed by an MLP for classification.

\begin{table*}[t]
\centering
\caption{Timing of HT detection per sample and training models.}
\vspace{-0.2cm}
\label{tab:timing}
\begin{tabular}{c|c|c|c|c|c|c|c|c|c|l|c|c|}
\cline{2-13}
\multicolumn{1}{l|}{} & \multicolumn{6}{c|}{\textbf{HT detection in RTL}} & \multicolumn{6}{c|}{\textbf{HT detection in netlist}} \\ \hline
\multicolumn{1}{|c|}{Benchmark} & AES & RS232 & PIC & DES & RC5 & \textbf{Ave.} & AES & RS232 & PIC & DES & RC5 & \textbf{Ave.} \\ \hline
\multicolumn{1}{|c|}{\begin{tabular}[c]{@{}c@{}}Detection\\ time(s)\end{tabular}} & 0.0268 & 0.0169 & 0.0182 & 0.0222 & 0.0215 & \textbf{0.0211} & 14.07 & 13.70 & 13.40 & 14.07 & 13.37 & \textbf{13.72} \\ \hline
\multicolumn{1}{|c|}{\begin{tabular}[c]{@{}c@{}}Training\\ time (s)\end{tabular}} & 220.37 & 251.18 & 274.18 & 261.71 & 284.53 & \textbf{258.39} & 1796 & 5449 & 5346 & 5164 & 4215 & \textbf{4394} \\ \hline
\end{tabular}
\vspace{-0.5cm}
\end{table*}
\section{Evaluation}
In this section, we explain our dataset and evaluation, report the results, and compare our proposed method with state of the art.

\subsection{Dataset}

We create a dataset of RTL codes and gate-level netlists based on the benchmarks in Trusthub \cite{trusthub1} which contain 34 varied types of HTs inserted in 3 base circuits: AES, PIC, and RS232. To expand our dataset, we extract the HTs design and use them as additional HT data instances. Moreover, we integrate some of the extracted HTs into new HT-free circuits, DES and RC5, which increases the number of HT-infested samples. To balance the dataset and increase the number of HT-free samples, we add different variations of open-source HT-free circuits in addition to new ones, including DET, RC6, SPI, SYN-SRAM, VGA, and XTEA to the dataset resulting in a dataset of 132 combined instances of 47 HT-Free and 85 HT-infested RTL codes. We also synthesized a corresponding generic gate-level netlist for each RTL code, which doubles the number of data instances. 
In our dataset, the number of nodes in netlist DFG of the base circuits on average is as follows: 109130 nodes in AES, 41649 nodes in DES, 107675 nodes in RC5, 10628 nodes in PIC, and 1307 nodes in RS232.

\begin{figure}[b]
\centering
\vspace{-0.5cm}
\includegraphics[width=0.45\textwidth]{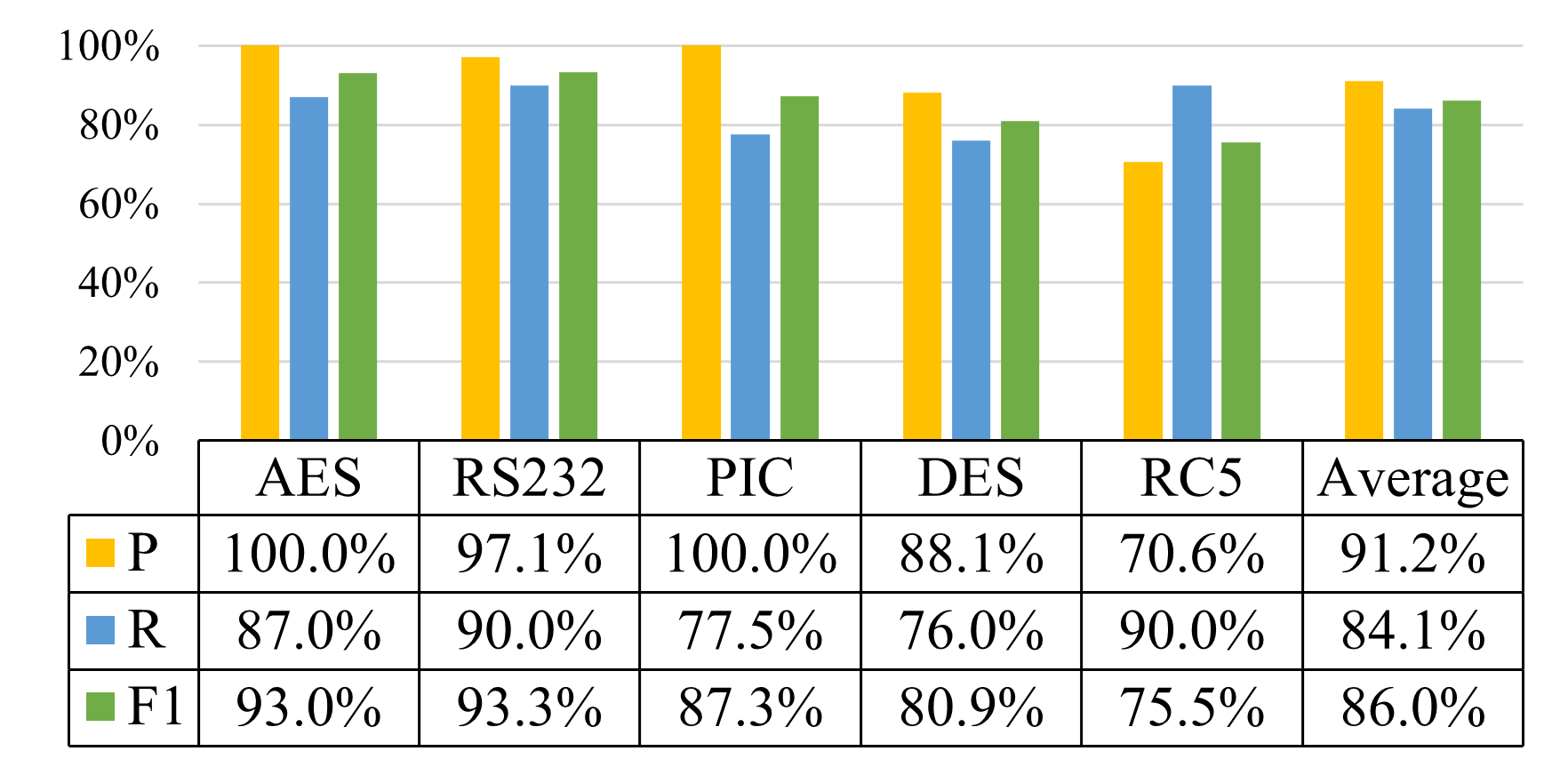}
\vspace{-0.3cm}
\caption{The performance of our method in gate-level netlist HT detection.}
\label{fig:netlist}
\end{figure}
\begin{figure}[b]
\centering
\vspace{-0.5cm}
\includegraphics[width=0.45\textwidth]{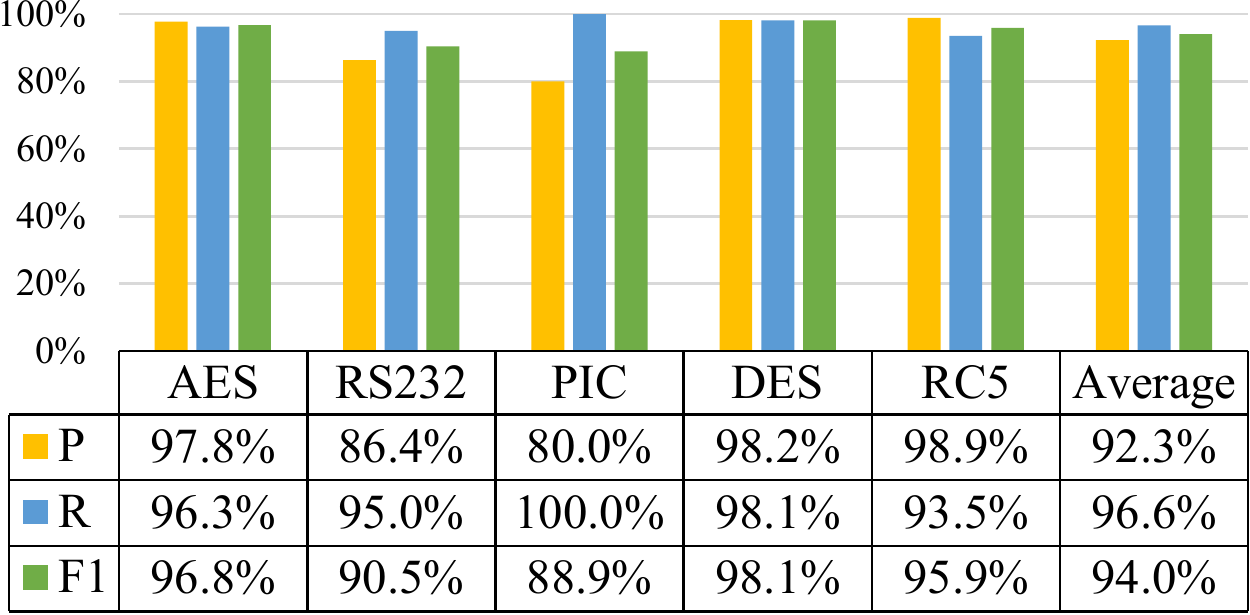}
\vspace{-0.3cm}
\caption{The performance of our method in RTL HT detection.}
\label{fig:RTL}
\end{figure}
\subsection{HT Detection Results}

We evaluate our method performance through an unknown HT detection scenario. To create this scenario, we apply the leave-one-out method, where we leave out one of the base circuits' HT-free and HT-infected benchmarks for testing and train the model on the rest of the benchmarks. This setup satisfies the claims of golden chip-free and unknown HT detection, as both golden HT-free versions of the circuit under test and the injected HT circuit are unknown to the model. We repeat this process 20 times for each base circuit and count the True Positive (TP), False Negative (FN), and False Positive (FP) to measure the evaluation metrics, Precision (P), Recall (R), or True Positive Rate, and $F_{\beta}~\-score$ as follows:

\vspace{+0.6em}
$P=\frac{TP}{TP + FP} , R=\frac{TP}{TP + FN}, F_{\beta}\-score=\frac{(1+\beta^2)*P*R}{\beta^2*P + R}$
\vspace{+0.6em}

Recall measures how well our model can detect HT-infested samples. This metric is intuitive, but it is not sufficient for evaluation, especially in an unbalanced dataset. It is also essential to have a low false-positive value which is the number of HT-free samples incorrectly classified as HT. Thus, precision is necessary as it measures the number of correctly classified HT samples over all samples classified as HT-infested, including the false positives. The $F_{\beta}~\-score$ is the weighted average of precision and recall that better presents the overall performance.

Figures \ref{fig:RTL} and \ref{fig:netlist} demonstrate the performance of our models for HT detection in RTL and netlist. Higher performance in RTL HT detection is justified by the intuition that RTL code is the behavioral description of the circuit, and the high level of abstraction assists with the learning process and facilitates the detection process. Moreover, the DFG generated from RTL is considerably smaller, and it is easier for the model to capture the key feature of the design. DFGs of netlists are more complicated and detailed, which impedes the learning process and requires a more complicated model. The complexity of the model can be adjusted by changing the number of hidden units and layers.

On the other hand, our models perform better for AES, RC5, and DES compared to RS232 and PIC because the former circuits are all encryption cores and the model has more data instances to learn their behavior while the latter circuits are a communication protocol and a processor. The performance for RS232 and PIC can be enhanced by including more circuits similar to PIC and RS232 in the dataset.

There exist circuits in RTL and netlist that do not follow the same pattern. For example, HT detection in RC5 has higher accuracy in RTL compared to RS232, whereas the reverse relation is recorded in the netlist. To analyze this case, multiple influential factors should be considered, such as the complexity of the circuit, the size of circuit DFG after optimization, the number of similar data samples in the dataset, etc. The combination of these factors has created such performance patterns, and by concentrating on a single aspect, the results may seem sporadic. In the case of RS232 and RC5 circuits, RS232 is a relatively simple circuit, and its netlist DFG is small and intuitive, which can be why high HT detection performance in RS232 is maintained, even in the netlist.

\subsection{Effect of Attention Mechanism}

\begin{figure}[t]
\centering
\includegraphics[width=0.45\textwidth]{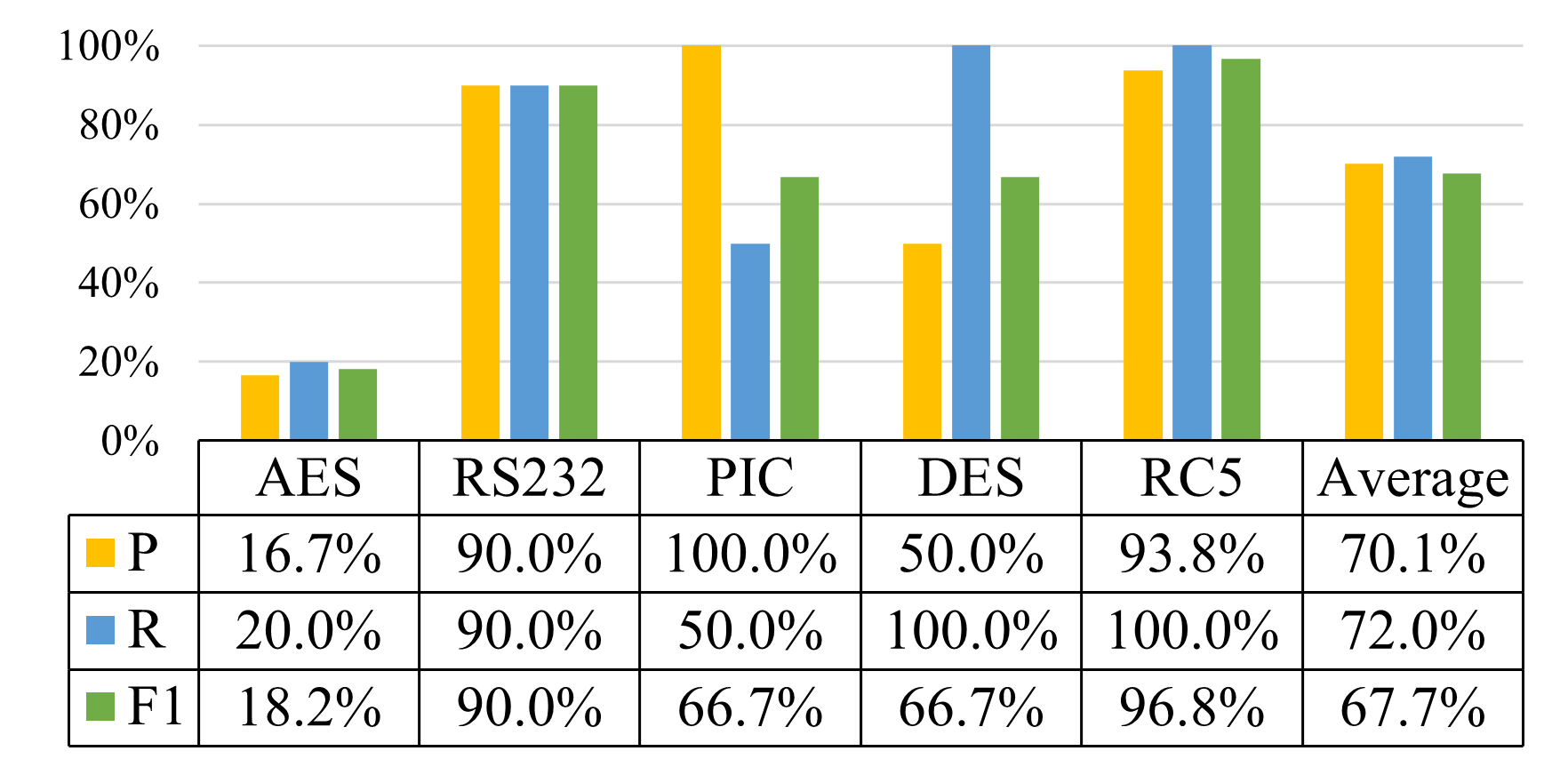}
\vspace{-0.2cm}
\caption{The performance of our model without pooling layer.}
\label{fig:pooling}
\vspace{-0.5cm}
\end{figure}

In this section, we investigate the effects of our attention mechanism, the pooling layer, on the performance of the model. In this case study, we keep the hyperparameters and architecture of the GNN model for HT detection in netlist the same and only change the pooling ratio to 1, which means the  operation does not occur. The results of this experiment are depicted in Figure~\ref{fig:pooling}. As the results indicate, in the absence of the pooling layer, learning does not happen correctly for most circuits. For example, for DES circuit recall equals 1 and precision is 0.5, which means the model has not learned the distinction between HT-free and HT-infected circuits and has labeled all the test samples as HT-infected. This experiment highlights that the attention mechanism plays a vital role in the learning process since it pushes the model to concentrate on the critical parts of the graph.

\begin{figure}[]
\centering
\includegraphics[width=0.45\textwidth]{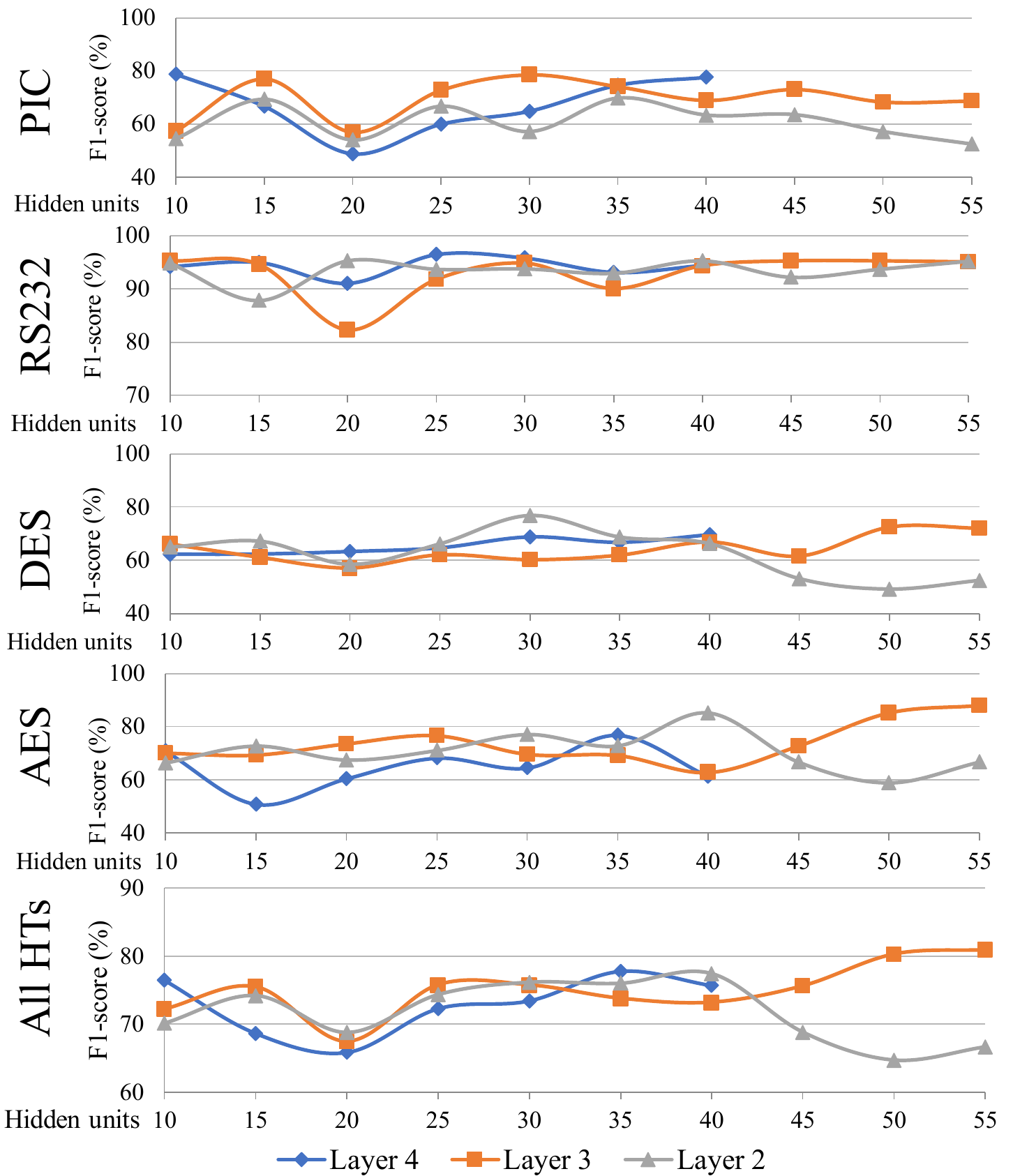}
\vspace{-0.2cm}
\caption{The model performance under different settings for number of convolutional layers and hidden units in each layer.}
\label{fig:netlist_layers}
\vspace{-0.5cm}
\end{figure}
\subsection{GNN Hyperparameters}  \label{sec:hyperparameters}

We develop two GNN models for HT detection in RTL and netlist. We determine hyperparameters of each model based on our intuition about the unique characteristics of our application, previous knowledge developed by common potent GNN architectures in the literature, and GNN design space exploration. The intuition toward the application usually helps with better configuration initialization and faster converges toward the optimal architecture. After testing a setting, we analyze the performance and decide what parameters to modify. For example, if we notice under-fitting, we increase the model complexity and computational power.

In the experiments on RTL, we use 2 GCN layers with 200 hidden units for each layer.
For the graph pooling layer, we use the pooling ratio of 0.8 to perform \textit{top-k filtering}. 
For \textbf{READOUT}, we use \textit{max-pooling} for aggregating node embeddings of each graph.
Our model uses 1 MLP layer to reduce the number of hidden units from 200 to 2 used in $\mathbf{h}_{G}$ for predicting the result of HT detection. In training, we append a dropout layer with a rate of 0.5 after each GCN layer. We train the model for 200 epochs using the batch gradient descent algorithm with batch size 4 and a learning rate of 0.001.

The gate-level netlists tend to produce considerably larger DFGs compared to RTL codes, even after optimization and node reduction steps. Thus, the netlist HT detection model requires a different configuration than the RTL model with an optimized structure to be easily trained using GPUs. 
In our constrained computation platform of a GPU with 14GB RAM, we limit the hidden units to 75 hidden units for 2-Layer networks, 58 for 3-layer, and 40 for 4-layer. We performed random search and grid search over a range of hidden layers, hidden units, pool ratio, and batch size. We found the average F1-score across all configurations for each layer is shown in Figure \ref{fig:netlist_layers}. According to the figure, the 3-layer network has the best average F1-score compared to the 2-layer and 4-layer networks. Fixing our model to a 3-layer network, we then find the best-hidden unit, pool ratio, and batch size by comparing the F1-score for all pairs of combinations across all 5 golden reference-free benchmarks. We found the best configuration for the gate-level netlist model to be a 3-layer network with 55 hidden units and a pool ratio of 0.6 for performing \textit{sagpool}, or self-attention graph pooling. The \textbf{READOUT} phase and the MLP are the same as the RTL model, and we train this model for 200 epochs using the batch gradient descent algorithm with batch size 2 and a learning rate of 0.001.
\subsection{Timing}

We train and test the RTL model on an NVIDIA GeForce GTX 1080 graphic card with 8GB memory and the gate-level netlist model on a V100 GPU with 16GB memory. The timing results for the base circuits are summarized in Table \ref{tab:timing}. Although training can be time-consuming, it occurs once, and the trained model finds the HT very fast in 21.1ms on average for RTL and 13.72 seconds for gate-level netlist which are both comparable to other works.

\begin{table*}[ht]
\centering
\caption{Comparing the performance of our method with the state-of-art methods for HT detection in 3PIP. }
\label{tab:compare}
\begin{tabular}{|C{0.9cm}|L{4cm}|C{1.2cm}|C{0.8cm}|C{0.8cm}|C{1.6cm}|C{1.2cm}|C{1.4cm}|C{2.6cm}|}
\hline
\textbf{Paper (year)}                 & \centering\textbf{Technique category - Method} & \textbf{Precision} & \textbf{Recall}  & \textbf{Time (s)} & \textbf{Automated feature extraction} & \textbf{Golden reference free} & \textbf{Unknown Trojan detection} & \textbf{HT detection scope is limited to}                                 
\\ \hline
Ours (2021)                                & ML - Graph neural network on RTL graph           & 92\%               & 97\%         & 0.026    & \cmark     & \cmark    & \cmark    & -                                                   
\\ \hline
Ours (2021)                                & ML - Graph neural network on netlist graph           & 91\%               & 84\%         & 13.72    & \cmark     & \cmark    & \cmark    & -                                                   
\\ \hline
\cite{TrojanAIS} (2018)               & ML - Artificial immune system       & 87\%               & 85\%         & NA       & \xmark     & \cmark    & \cmark    & -                                                   
\\ \hline
\cite{boostAST} (2019)                 & ML - Gradient boosting algorithm    & NA                 & 100\%        & 1.36     & \xmark     & \xmark    & \cmark    & -                                                   
\\ \hline
\cite{hasegawa2017hardware} (2017)    & ML - Multi-layer neural networks    & NA                 & 90\%         & NA       & \xmark     & \xmark    & \cmark    & -                                                   
\\ \hline  
\cite{demrozi2017exploiting} (2017)   & ML,GM - Subgraph isomorphism        & NA                 & 100\%        & 1.15     & \xmark     & \cmark    & \xmark    & HTs in the library    
\\ \hline
\cite{fyrbiak2020graph} (2020)        & GM - Graph similarity               & NA                 & NA           & NA       & \xmark     & \cmark    & \xmark    & -   
\\ \hline
\cite{piccolboni2017efficient} (2017) & GM - Subgraph matching              & NA                 & 100\%        & 5.02     & \xmark     & \cmark    & \xmark    & HTs in the library   
\\ \hline
\cite{islam2019socio} (2019)          & CA - Socio-network analysis         & 98\%               & 98\%         & NA       & \xmark     & \cmark    & \xmark    & combinational HTs
\\ \hline
\cite{nahiyan2017hardware} (2017)     & FV - Information flow analysis      & NA                 & 100\%        & 292.85   & \xmark     & \cmark    & \xmark    & -                                                   
\\ \hline
\cite{rajendran2016formal} (2016)     & FV - Model checking                 & NA                 & 100\%        & 96.13    & \xmark     & \cmark    & \xmark    & HTs that leak data             
\\ \hline
\end{tabular}
\vspace{-0.5cm}
\end{table*}
\subsection{Comparison with State of the Art}
In this section, we compare leading pre-silicon HT detection techniques, in terms of quantitative and qualitative metrics as well as in a case study.

\subsubsection{\textbf{Qualitative comparison}}
We compare our models against other works using 3 essential qualitative metrics for HT detection; i) golden reference-free, ii) unknown HT detection, and iii) automated process. Designing an effective HT detection method without white-box knowledge or access to a trusted reference design is very challenging as both are not readily available. Traditional algorithmic approaches that utilize heuristics can satisfy the golden reference-free condition, whereas ML techniques naturally rely on datasets containing both HT-free and HT-infested designs to train and classify test circuits. The next important metric is the ability to detect unknown HTs. In this regard, ML-based approaches have more success as the features of existing circuits are learned and used to classify similar circuits or unseen circuits. Techniques such as \cite{piccolboni2017efficient, fyrbiak2020graph} and \cite{rajendran2016formal, nahiyan2017hardware} limit their detection scope to HTs that exist in the method's library or only to security properties that are explicitly defined. To the best of our knowledge, all the existing approaches require feature engineering or manual property definition. However, our approach is capable of HT detection with an automated feature extraction process, which removes the need for a manual feature or property extraction. Thus, our models can easily expand to various types of circuits and can scale to industrial-level IP designs through retraining if a new type of HT is discovered.

\subsubsection{\textbf{Quantitative comparison}}
For quantitative comparison, we analyze the precision, recall, and timing metrics. For timing, we compare the average detection time for the AES benchmarks due to being the largest benchmarks in our dataset in terms of graph size.

It should be noted that the exact comparison between the reported timings is not possible as the computing platforms differ on a paper-to-paper basis. However, the relative difference between the timings shows that our models are faster than if not comparable to others. The timing of algorithmic methods (FV, CA, and GM) highly depends on the hardware design complexity. Their memory usage and detection time drastically grow for large designs which can cause timeout or memory shortage problems. On the other hand, the circuit's complexity does not have a notable effect on the detection time of the proposed RTL method according to Table \ref{tab:compare}, which makes it scalable for large designs. Our gate-level netlist detection time is consistent across benchmarks of varying complexity which shows that it is also not greatly affected by the circuit complexity.

For recall and precision, both our RTL and gate-level netlist have comparable results. Although it is challenging to compare against algorithm methods since most papers only report a list of known HT benchmarks that their models can successfully detect while we test our model with the unknown HTs scenario, which is mostly not detectable by those algorithms. Furthermore, FP and TP for computing precision and the performance of the methods on HT-free samples are not available in some papers to compare.

\subsubsection{\textbf{Case study}}

We analyze the ML, GM, and FV techniques in a case study and investigate if the models can detect 6 different types of HTs. In Table \ref{tab:test-case}, the first top 3 HT benchmarks are known to the Method Under Test (MUT) and exist in its library of HTs whereas the other 3 HTs are unknown.  \cmark \cmark indicates that the MUT can detect the HT and it is explicitly reported in its paper. \cmark\textbackslash\xmark  shows that this case is not tested in the paper but it is supposed to detect\textbackslash not to detect the HT according to authors' claims and assumptions. The results indicate that GM methods rely on the HT library and memorize them and consequently, cannot recognize the HTs out of the library. FV methods depend on the predefined properties for HTs and cannot identify new types of HTs. On the other hand, our method is an ML-based method that learns the HT behaviors and can pinpoint different types of HT, even the unknown ones.
\begin{table*}[t]
\centering
\caption{Comparing HT detection methods in a case study.}
\label{tab:test-case}
\begin{tabular}{|c|c|c|c|c|c|c|c|c|}
\hline
\textbf{Benchmark}  & \textbf{Victim Circuit} & \textbf{Trigger} & \textbf{Payload}           &\textbf{Ours (ML)} & \textbf{\cite{fyrbiak2020graph} (GM)} & \textbf{\cite{piccolboni2017efficient} (GM)} & \textbf{\cite{rajendran2016formal} (FV)} & \textbf{\cite{nahiyan2017hardware} (FV)} \\ \hline
AES-T900    & AES encryption core & Time bomb  &   Leak data      & \cmark\cmark       & \cmark                       & \cmark \cmark                       & \cmark \cmark                   & \cmark \cmark                   \\ \hline
RS232-T500  & UART serial communication & Time bomb  &   Deny service & \cmark\cmark       & \cmark                       & \cmark \cmark                       & \xmark                          & \cmark \cmark                   \\ \hline
AES-T1900     & AES encryption core & Time bomb  &   Degrade chip      & \cmark\cmark       & \cmark                       & \cmark \cmark                       & \xmark                          & \xmark                          \\ \hline
AES-T2000     & AES encryption core&  Cheat code  &  Leak data      & \cmark\cmark       & \xmark                       & \xmark                              & \cmark \cmark                   & \cmark \cmark                   \\ \hline
RS232-T700   & UART serial communication & Cheat Code & Deny service & \cmark\cmark       & \xmark                       & \xmark                              & \xmark                          & \cmark                          \\ \hline
AES-T1800     & AES encryption core & Cheat code  & Degrade chip      & \cmark\cmark       & \xmark                       & \xmark                              & \xmark                          & \xmark                          \\ \hline
\end{tabular}
\end{table*}
\section{Related Works and Backgrounds}\label{sec:related-works}

\subsection{Hardware Trojan}
An HT consists of two fundamental parts: payload and trigger. The payload is the implementation of the malicious behavior of the HT. This malicious behavior could lead to information leaks such as the secret key of a cryptographic core which would enable unrestricted access to sensitive data. It could change the functionality of the circuit to sabotage or cause harm. It can deny the service of certain functionalities of the chip or degrade the performance by increasing power consumption. The trigger is an optional circuit that monitors various signals or events in the base circuit and activates the payload when a specific signal or event is observed. HTs without triggers are usually always-on Trojans. HTs with triggers in the context of digital circuits, can either be combinational or sequential. Combinational triggers activate upon a specific set of signal inputs. Sequential triggers rely on a specific sequence of signals or events. A sequential trigger could be based on a counter reaching a specific value or when a signal pattern repeats over a certain duration. A sequential trigger is more difficult to activate compared to a combinational trigger due to the vast number of states that are required to be checked. In our paper, we classify the triggers of HTs into three main categories; i) time bomb, ii)  cheat code,  and iv) always on. The time bomb trigger is activated after numerous clock cycles and the cheat code trigger depends on one or a sequence of specific inputs in the base circuit.

HTs are designed with malicious intent and to stay undetected. HT designers can reduce the likelihood of their HTs from being detected by using a common technique that involves placing them in rare internal nodes with exceedingly low activation probabilities. Rare internal nodes are difficult to detect because they are often outside the functional context of the intended design, and they often go undetected by traditional test methods. Therefore, many HT detection methods take into account the rare internal node characteristic of HTs as part of their detection schemes.
\subsection{Countermeasures for Hardware Trojan}
The majority of the pre-silicon HT detection techniques in the literature fall into four main categories, described below:
\subsubsection{\textbf{Test Pattern Generation}} 

Traditional test pattern generation has been widely used in both pre-silicon and post-silicon manufacturing stages. The idea is to create an extensive set of test vectors with varying input logic combinations to validate a circuit's output against the intended design output. The intended design output must come from a trusted golden reference chip of the original circuit design. These test patterns are typically generated based on the system specifications of the design which describes the functionalities and behaviors of the circuit. However, due to its stealthy nature, HTs remain inactive and well-hidden during simulation and testing. Many HTs can bypass detection from these test vectors by simply not modifying the functionalities of the original circuit. More importantly, most HTs only activate under particular rare events. This makes it difficult for traditional test generation methods to trigger HTs that use internal node triggers with extremely low activation probability.

To tackle these issues, the authors in \cite{mero,saha2015improved} use automatic test pattern generation (ATPG) to create effective test vectors by identifying rare input logic at internal nodes to increase the probability of triggering the HT. However, even with this strategy, the scale of modern designs makes it possible for HT designers to create extremely low activation triggers that could evade detection by combining multiple rare signals and having the trigger conditions occur over multiple steps. \cite{TARMAC} proposed a new test generation method by mapping the trigger activation problem to a clique cover problem which can detect extremely rare triggers, but is shown to have unstable coverage performance. A recent study employs reinforcement learning techniques in conjunction with rare node excitation, as well as controllability and observability analysis to generate test vectors with improved trigger coverage and test generation time. 
Most methods above assume a full-scan chain design which simplifies ATPG by converting sequential elements into combinational elements with scan flip flops. However, not all designs have full-scan chains due to design constraints such as power, area, and additional hardware components required \cite{BHUNIAxScanChain}. Therefore, a partial-scan chain design has been adopted. \cite{ATPG&Model} tackles the challenges of partial-scan designs by combining ATPG with model checking for more efficient test vector generation and improved HT coverage. Still, there is no guarantee of success in the test pattern generation approach, most test generation techniques are time-consuming due to their iterative nature. More importantly, in order to achieve full coverage, test pattern generation would need to generate and test all the possible cases, but given the scale of modern ICs, it is simply infeasible since the state space for test vectors grows exponentially to the number of rare input signals.
\subsubsection{\textbf{Formal Verification (FV)}}

Formal verification is a mathematical proof-checking technique that relies on the security and trust policies defined in the system-level specification. It verifies the integrity of the design using common verification methods such as property checking, equivalence checking, and model checking. In order to apply formal verification, the 3PIP design must first be converted from an HDL language like Verilog to an equivalent model in the proof-checking format using a formal language like Gallina, a tactic language developed as part of the well-known theorem prover called Coq. The 3PIP is delivered with a separate code representing the 3PIP but written in a formal language and used for proof-checking; this code is known as proof-carrying code (PCC). PCC can either be provided by the 3PIP vendor directly or they can provide a Soft IP that can be converted into a proof-checking format.

Applying FV methods, \cite{subramanyan2014formal} formulate taint-propagation properties that verify the data flow between signals in a design to identify unintentional design bugs. HTs, however, are intentional by nature, so the criteria for bug detection do not directly apply to HT detection. A similar approach is able to detect information leakage \cite{rajendran2016formal} and malicious modification to registers \cite{rajendran2015detecting} by applying information leakage and register modification as criteria for the security properties. While using formal verification on 3PIP proves the predefined security properties, its detection scope is limited to the properties stated in the system-level specification. Only specific types of HTs can be detected because the properties are insufficient to cover all the various types of malicious behaviors that HTs can exhibit. \cite{rajendran2015detecting}, for example, defines ``no-critical-data-corruption" which can only detect data-corrupting HTs. \cite{rajendran2016formal} modifies the security criteria from data corruption to data leakage. Both approaches employ model checking, which does not scale to large designs because model checking is NP complex and suffers from state explosion. \cite{Guo2016ScalableST} combines theorem proving with model checking to overcome the state explosion issues of model checking, however still suffers from the limited scope issue. Information flow tracking has been used to model security properties that can provide wider coverage of HTs \cite{nahiyan2017hardware}. FV does not depend on HT trigger conditions for detection, so it does not suffer from the issue of not being able to detect HTs due to low trigger probability. However, it is still possible for 3PIP vendors to intelligently manipulate the proofs and security properties to evade FV.
\subsubsection{\textbf{Code Analysis (CA)}}
Code analysis or code coverage is a technique that analyzes the execution of the RTL or gate-level netlist code. To verify the 3PIP for trust, code analysis uses metrics such as line, statement, toggle, and finite state machine (FSM) coverage and compares against the design specification to ascertain the suspicious signals that imitate the HT. These coverage metrics respectively check which lines and statements are executed, what signals are being switched in the gate-level netlist, and which states are reached during execution. If anything less than 100\% coverage is reported, then the 3PIP design is considered to be HT infested.

Using CA, \cite{Nahiyan2016AVFSMAF} extracts state transition graphs from gate-level netlist and reports state transitions that are vulnerable to HT injection. This approach burdens the designer with manual analysis of the suspicious regions to identify the possible HT. It is also limited to the design's combinational logic. FANCI \cite{waksman2013fanci} proposed a control value metric that measures the degree of influence a given input has on the operation and output of a circuit. It looks for nets in the HDL code with very low control values and marks them as suspicious nets. VeriTrust \cite{zhang2015veritrust} returns the nets that are not driven by functional inputs as potential triggers for an HT. DeTrust \cite{zhang2014detrust}, on the other hand, proposes an attack that exploits the vulnerabilities of FANCI and VeriTrust by modifying the HTs with stealthy implicit triggers. Trigger logic for these HTs is distributed over multiple stages with a combination of sequential and combinational logic alongside logic that is part of the intended design, making them much more challenging to detect. In order to execute/cover all the lines of the code, code analysis requires an effective test bench to cover all the execution scenarios of the design. This causes the same problems as test pattern generation in that the number of test patterns needed for higher coverage scales poorly with larger designs and verification time increases to an infeasible amount. It has also been shown that even 100\% coverage does not guarantee that a 3PIP is HT-free \cite{Jou1999CoverageAT}.
\subsubsection{\textbf{Machine Learning (ML)}}
Machine learning is a powerful technique that has received a lot of attention lately and has revolutionized different fields of study \cite{yasaei2020iot, ashrafiamiri2020r2ad}. The majority of ML-based works for HT detection depend on side-channel signal analysis \cite{htnet, htm, ashrafiamiri2018towards}, which delays the detection until the post-silicon stages. More recent approaches have focused on extracting various features during the design phase of the circuit to classify the HT-free and HT-infested 3PIPs. For example, the gradient boosting algorithm uses the Abstract Syntax Tree (AST) of RTL code \cite{boostAST}, the multilayer neural network uses Trojan-net features of gate-level netlist \cite{hasegawa2017hardware}, the artificial immune system uses DFG and Control Flow Graph (CFG) of RTL code \cite{TrojanAIS}, and the probabilistic neural network uses CFG of RTL code \cite{demrozi2017exploiting}. Most of these models rely on an HT-free golden reference. However, a trusted reference is not guaranteed because the untrusted 3PIP vendors are the ones that provide the source code and specifications that may include hidden HTs as part of the golden reference. Moreover, reference-based methods may be inconclusive or too complex for exhaustive verification, especially for large designs. With classical ML models, the models' performance heavily depends on the quality of the selected features. This typically requires a lot of upfront resource investment in feature engineering and can be quite time-consuming. Additional feature engineering would also be required to account for newly developed and unknown HTs. We propose a scalable, golden reference-free model that leverages a potent ML technique, GNN, which performs automatic feature extraction and learns the circuit's behavior, both intended and malicious behaviors, to detect known and unknown HTs.
\subsection{Graph in Hardware Applications}
A graph is an intuitive representation of a hardware design. The vast network of gates in a circuit can be naturally represented by the interconnections of nodes as gates and edges as wires in a graph. Graph-related problems have been around for decades with firm roots in discrete mathematics and computer science. There is a rich pool of knowledge in graph theory and many well-established algorithms were developed to solve these graph problems. By transforming a hardware design into its graphical representation, we can apply the same concepts and algorithms in solving graph problems to solve hardware design problems.

Various graph representations of traditional EDA problems are shown in \cite{graphsinEDA}. For example, logical verification can be modeled using a rooted directed graph to represent the boolean function \cite{graph_boolean}. It is clear that many EDA problems can be converted into graph problems. However, many graph problems suffer from NP complexity and need to overcome scalability issues due to large modern IC designs. In more recent works, \cite{piccolboni2017efficient} propose analyzing the data/control flow graph of the circuit to locate the HTs. The authors create a library of HTs and use sub-graph matching algorithms to find the graphs of such known HTs in the graph of the 3PIP designs. Graph matching is an NP-complex problem that does not scale to large designs. To improve the accuracy and computation time, \cite{fyrbiak2020graph} introduces a new graph similarity heuristic tailored for hardware security. These methods can detect only the HTs that have the same graph representation as known HTs in their library, while attackers will design a diverse set of HTs. Recently, GNN has been deployed for reverse engineering to assist with malicious logic detection. In this direction, ReIGNN \cite{ReIGNN} has combined GNN with structural analysis to classify registers in a netlist. GNN-RE \cite{alrahis2021gnn} leverages GNN to identify the boundaries of modules in a flattened netlist and classify the functionality of sub-circuits.

Recently GNN models have grabbed attention in EDA and hardware security \cite{hw2vec, gnn4ip, gnn4tj}. GNN is deep learning that operates on the graph and has the advantages of the machine learning methods but can be applied to non-Euclidean data. Our intuition behind using GNN is that graphs are an intrinsic representation of a hardware design.  Representing the hardware design in the form of DFGs allows us to capture the behavior between signals, sequential elements, and combinational elements in the circuit. Through the DFGs, the GNN will learn to distinguish normal circuits from the circuits that contain malicious functions.
\section{Discussion}

The size of the graph dataset plays an important role in determining how well our models can learn to generalize with unknown circuit designs. The more nodes in the graph, the more difficult it is for the learning process. The difference in nodes between the gate-level netlist representation versus the RTL representation is several orders of magnitudes larger. Due to the memory limitations, the complexity of our model was also limited to a specific range of layers and hidden units. Future works should look at how to further reduce the graph sizes for the gate-level netlist or to use a different type of graph representation for the hardware circuit.
\section{Conclusion}
This paper proposes a novel golden reference-free approach to find unknown HT in both RTL codes and gate-level netlists. We generate DFGs for both RTL and netlist codes and employ the GNN to construct two models that infer the presence of HT from the generated graphs in RTL and gate-level netlist. Our method automatically extracts the features of graphs and learns the behavior of the hardware design. Our model is trained and tested on a DFG dataset created by expanding the Trustub benchmarks. The RTL results indicate that the proposed method discovers HT with 97\% recall and 94\% F1-score very fast in 21.1ms. The gate-level netlist results indicate an 84\% recall and 86\% F1-score with an average detection time of 13.72 seconds.
\section*{Acknowledgment}
This research was supported by the Office of Naval Research (ONR), award number N00014-17-1-2499. Any opinions, findings, conclusions, or recommendations expressed in this material are those of the authors and do not necessarily reflect the views of our funding agencies.

\ifCLASSOPTIONcaptionsoff
  \newpage
\fi



%

\bibliographystyle{IEEEtran}
\bibliography{bibliography}

\begin{thebibliography}{10}
\providecommand{\url}[1]{#1}
\csname url@samestyle\endcsname
\providecommand{\newblock}{\relax}
\providecommand{\bibinfo}[2]{#2}
\providecommand{\BIBentrySTDinterwordspacing}{\spaceskip=0pt\relax}
\providecommand{\BIBentryALTinterwordstretchfactor}{4}
\providecommand{\BIBentryALTinterwordspacing}{\spaceskip=\fontdimen2\font plus
\BIBentryALTinterwordstretchfactor\fontdimen3\font minus
  \fontdimen4\font\relax}
\providecommand{\BIBforeignlanguage}[2]{{%
\expandafter\ifx\csname l@#1\endcsname\relax
\typeout{** WARNING: IEEEtran.bst: No hyphenation pattern has been}%
\typeout{** loaded for the language `#1'. Using the pattern for}%
\typeout{** the default language instead.}%
\else
\language=\csname l@#1\endcsname
\fi
#2}}
\providecommand{\BIBdecl}{\relax}
\BIBdecl

\bibitem{trusthub1}
B.~Shakya, M.~He, H.~Salmani, D.~Forte, S.~Bhunia, and M.~Tehranipoor,
  ``Benchmarking of hardware trojans and maliciously affected circuits,''
  \emph{Journal of Hardware and Systems Security}, 2017.

\bibitem{syrianRadar}
S.~Adee, ``The hunt for the kill switch,'' in \emph{IEEE Spectrum}, 2008.

\bibitem{MilitaryChipBackdoor}
S.~Skorobogatov and C.~Woods, ``Breakthrough silicon scanning discovers
  backdoor in military chip,'' in \emph{Cryptographic Hardware and Embedded
  Systems -- CHES 2012}, E.~Prouff and P.~Schaumont, Eds.\hskip 1em plus 0.5em
  minus 0.4em\relax Springer Berlin Heidelberg, 2012.

\bibitem{MarketAnalysis2020-2025}
\BIBentryALTinterwordspacing
``Semiconductor intellectual property (ip) market with covid-19 impact
  analysis,'' accessed: 2021-04-20. [Online]. Available:
  \url{https://www.researchandmarkets.com/reports/5185336/
  semiconductor-intellectual-property-ip-market\#pos-0}
\BIBentrySTDinterwordspacing

\bibitem{cadence2019tortuga}
N.~Fern and S.~Carlson, ``A complete system-level security verification
  methodology. (white paper from cadence and tortuga logic),'' 2019.

\bibitem{UCIxhicks}
M.~Hicks, M.~Finnicum, S.~T. King, M.~M.~K. Martin, and J.~M. Smith,
  ``Overcoming an untrusted computing base: Detecting and removing malicious
  hardware automatically,'' in \emph{2010 IEEE Symposium on Security and
  Privacy}, 2010.

\bibitem{SMC}
C.~Sturton, M.~Hicks, D.~Wagner, and S.~T. King, ``Defeating uci: Building
  stealthy and malicious hardware,'' in \emph{Proceedings of the 2011 IEEE
  Symposium on Security and Privacy}.\hskip 1em plus 0.5em minus 0.4em\relax
  IEEE Computer Society, 2011.

\bibitem{waksman2013fanci}
A.~Waksman~et al., ``Fanci: identification of stealthy malicious logic using
  boolean functional analysis,'' in \emph{ACM SIGSAC Conference on Computer and
  Communications Security}, 2013.

\bibitem{zhang2014detrust}
J.~Zhang, F.~Yuan, and Q.~Xu, ``Detrust: Defeating hardware trust verification
  with stealthy implicitly-triggered hardware trojans.''\hskip 1em plus 0.5em
  minus 0.4em\relax Association for Computing Machinery, 2014.

\bibitem{mero}
R.~S. Chakraborty, S.~Wolff, Francisand~Paul, C.~Papachristou, and S.~Bhunia,
  ``Mero: A statistical approach for hardware trojan detection,'' in
  \emph{Cryptographic Hardware and Embedded Systems - CHES 2009}, 2009.

\bibitem{saha2015improved}
S.~Saha~et al., ``Improved test pattern generation for hardware trojan
  detection using genetic algorithm and boolean satisfiability,'' in
  \emph{International Workshop on Cryptographic Hardware and Embedded Systems},
  2015.

\bibitem{rajendran2015detecting}
J.~Rajendran~et al., ``Detecting malicious modifications of data in third-party
  intellectual property cores,'' in \emph{ACM/IEEE Design Automation Conference
  (DAC)}, 2015.

\bibitem{rajendran2016formal}
------, ``Formal security verification of third party intellectual property
  cores for information leakage,'' in \emph{International Conference on VLSI
  Design and Embedded Systems (VLSID)}, 2016.

\bibitem{fyrbiak2020graph}
M.~Fyrbiak~et al., ``Graph similarity and its applications to hardware
  security,'' \emph{IEEE Tran. on Computers}, 2020.

\bibitem{piccolboni2017efficient}
L.~Piccolboni~et al., ``Efficient control-flow subgraph matching for detecting
  hardware trojans in rtl models,'' \emph{ACM Tran. on Embedded Computing
  Systems (TECS)}, 2017.

\bibitem{MultilevelFASTrust2018}
X.~Chen, Q.~Liu, S.~Yao, J.~Wang, Q.~Xu, Y.~Wang, Y.~Liu, and H.~Yang,
  ``Hardware trojan detection in third-party digital intellectual property
  cores by multilevel feature analysis,'' \emph{IEEE Transactions on
  Computer-Aided Design of Integrated Circuits and Systems}, 2018.

\bibitem{kipf2016semi}
T.~N. Kipf and M.~Welling, ``Semi-supervised classification with graph
  convolutional networks,'' \emph{arXiv preprint arXiv:1609.02907}, 2016.

\bibitem{liu2020efficient}
S.~Liu, J.~H. Park, and S.~Yoo, ``Efficient and effective graph convolution
  networks,'' in \emph{Proceedings of the 2020 SIAM International Conference on
  Data Mining}.\hskip 1em plus 0.5em minus 0.4em\relax SIAM, 2020.

\bibitem{knyazev2019understanding}
B.~Knyazev~et al., ``Understanding attention and generalization in graph neural
  networks,'' in \emph{Advances in Neural Information Processing Systems
  (NeurIPS)}, 2019.

\bibitem{lee2019self}
J.~Lee~et al., ``Self-attention graph pooling,'' \emph{arXiv preprint
  arXiv:1904.08082}, 2019.

\bibitem{takamaeda2015pyverilog}
S.~Takamaeda-Yamazaki, ``Pyverilog: A python-based hardware design processing
  toolkit for verilog hdl,'' in \emph{International Symposium on Applied
  Reconfigurable Computing}, 2015.

\bibitem{Yosys}
C.~Wolf, ``Yosys open synthesis suite,'' \url{http://www.clifford.at/yosys/}.

\bibitem{TrojanAIS}
F.~Zareen and R.~Karam, ``Detecting rtl trojans using artificial immune systems
  and high level behavior classification,'' in \emph{Asian Hardware Oriented
  Security and Trust Symposium (AsianHOST)}, 2018.

\bibitem{boostAST}
T.~{Han et al.}, ``Hardware trojans detection at register transfer level based
  on machine learning,'' in \emph{Symposium on Circuits and Systems(ISCAS)},
  2019.

\bibitem{hasegawa2017hardware}
K.~Hasegawa~et al., ``Hardware trojans classification for gate-level netlists
  using multi-layer neural networks,'' in \emph{IEEE International Symposium on
  On-Line Testing and Robust System Design (IOLTS)}, 2017.

\bibitem{demrozi2017exploiting}
F.~Demrozi~et al., ``Exploiting sub-graph isomorphism and probabilistic neural
  networks for the detection of hardware trojans at rtl,'' in \emph{IEEE
  International High Level Design Validation and Test Workshop}, 2017.

\bibitem{islam2019socio}
S.~A. Islam~et al., ``Socio-network analysis of rtl designs for hardware trojan
  localization,'' in \emph{International Conference on Computer and Information
  Technology (ICCIT)}, 2019.

\bibitem{nahiyan2017hardware}
A.~Nahiyan~et al., ``Hardware trojan detection through information flow
  security verification,'' in \emph{IEEE International Test Conference (ITC)},
  2017.

\bibitem{TARMAC}
Y.~Lyu and P.~Mishra, ``Automated trigger activation by repeated maximal clique
  sampling,'' in \emph{2020 25th Asia and South Pacific Design Automation
  Conference (ASP-DAC)}, 2020.

\bibitem{BHUNIAxScanChain}
S.~Bhunia and M.~Tehranipoor, ``Chapter 3 - system on chip (soc) design and
  test,'' in \emph{Hardware Security}, 2019.

\bibitem{ATPG&Model}
J.~Cruz, F.~Farahmandi, A.~Ahmed, and P.~Mishra, ``Hardware trojan detection
  using atpg and model checking,'' in \emph{2018 31st International Conference
  on VLSI Design and 2018 17th International Conference on Embedded Systems
  (VLSID)}, 2018.

\bibitem{subramanyan2014formal}
P.~Subramanyan and D.~Arora, ``Formal verification of taint-propagation
  security properties in a commercial soc design,'' in \emph{Design, Automation
  \& Test in Europe Conference (DATE)}, 2014.

\bibitem{Guo2016ScalableST}
X.~Guo, R.~Dutta, P.~Mishra, and Y.~Jin, ``Scalable soc trust verification
  using integrated theorem proving and model checking,'' \emph{2016 IEEE
  International Symposium on Hardware Oriented Security and Trust (HOST)},
  2016.

\bibitem{Nahiyan2016AVFSMAF}
A.~Nahiyan, K.~Xiao, K.~Yang, Y.~Jin, D.~Forte, and M.~Tehranipoor, ``Avfsm: A
  framework for identifying and mitigating vulnerabilities in fsms,''
  \emph{2016 53nd ACM/EDAC/IEEE Design Automation Conference (DAC)}, 2016.

\bibitem{zhang2015veritrust}
J.~Zhang~et al., ``Veritrust: Verification for hardware trust,'' \emph{IEEE
  Tran. on Computer-Aided Design of Integrated Circuits and Systems}, 2015.

\bibitem{Jou1999CoverageAT}
J.-Y. Jou and C.~Liu, ``Coverage analysis techniques for hdl design
  validation,'' 1999.

\bibitem{yasaei2020iot}
R.~Yasaei, F.~Hernandez, and M.~A. Al~Faruque, ``Iot-cad: context-aware
  adaptive anomaly detection in iot systems through sensor association,'' in
  \emph{IEEE/ACM International Conference On Computer Aided Design (ICCAD)},
  2020.

\bibitem{ashrafiamiri2020r2ad}
M.~Ashrafiamiri, S.~M.~P. Dinakarrao, A.~H.~A. Zargari, M.~Seo, F.~Kurdahi, and
  H.~Homayoun, ``R2ad: Randomization and reconstructor-based adversarial
  defense for deep neural networks,'' in \emph{ACM/IEEE 2nd Workshop on Machine
  Learning for CAD (MLCAD)}, 2020.

\bibitem{htnet}
S.~Faezi, R.~Yasaei, and M.~A. Al~Faruque, ``Htnet: Transfer learning for
  golden chip-free hardware trojan detection,'' in \emph{IEEE Design,
  Automation \& Test in Europe Conference \& Exhibition (DATE)}, 2021.

\bibitem{htm}
S.~Faezi, R.~Yasaei, A.~Barua, and M.~A. Al~Faruque, ``Brain-inspired golden
  chip free hardware trojan detection,'' \emph{IEEE Transactions on Information
  Forensics and Security (TIFS)}, 2021.

\bibitem{ashrafiamiri2018towards}
M.~AshrafiAmiri, A.~H.~A. Zargari, S.~M.-H. Farzam, and S.~Bayat-Sarmadi,
  ``Towards side channel secure cyber-physical systems,'' in \emph{IEEE
  Real-Time and Embedded Systems and Technologies (RTEST)}, 2018.

\bibitem{graphsinEDA}
Y.~Ma, Z.~He, W.~Li, L.~Zhang, and B.~Yu, ``Understanding graphs in eda: From
  shallow to deep learning,'' ser. ISPD '20.\hskip 1em plus 0.5em minus
  0.4em\relax Association for Computing Machinery, 2020.

\bibitem{graph_boolean}
Bryant, ``Graph-based algorithms for boolean function manipulation,'' 1986.

\bibitem{ReIGNN}
S.~D. Chowdhury, K.~Yang, and P.~Nuzzo, ``Reignn: State register identification
  using graph neural networks for circuit reverse engineering,'' in \emph{2021
  IEEE/ACM International Conference On Computer Aided Design (ICCAD)}, 2021.

\bibitem{alrahis2021gnn}
L.~Alrahis, A.~Sengupta, J.~Knechtel, S.~Patnaik, H.~Saleh, B.~Mohammad,
  M.~Al-Qutayri, and O.~Sinanoglu, ``Gnn-re: Graph neural networks for reverse
  engineering of gate-level netlists,'' \emph{IEEE Transactions on
  Computer-Aided Design of Integrated Circuits and Systems}, 2021.

\bibitem{hw2vec}
S.-Y. Yu, R.~Yasaei, Q.~Zhou, T.~Nguyen, and M.~A. Al~Faruque, ``Hw2vec: A
  graph learning tool for automating hardware security,'' in \emph{IEEE
  International Symposium on Hardware Oriented Security and Trust (HOST'21)},
  2021.

\bibitem{gnn4ip}
R.~Yasaei, S.-Y. Yu, E.~K. Naeini, and M.~A.~A. Faruque, ``Gnn4ip: Graph neural
  network for hardware intellectual property piracy detection,'' in
  \emph{ACM/IEEE Design Automation Conference (DAC)}, 2021.

\bibitem{gnn4tj}
R.~Yasaei, S.-Y. Yu, and M.~A. Al~Faruque, ``Gnn4tj: Graph neural networks for
  hardware trojan detection at register transfer level,'' in \emph{IEEE Design,
  Automation \& Test in Europe Conference \& Exhibition (DATE)}, 2021.

\end{thebibliography}


\section{Biography Section}
\vspace{-1cm}
\begin{IEEEbiography}[{\includegraphics[width=1in,height=1.25in,clip,keepaspectratio]{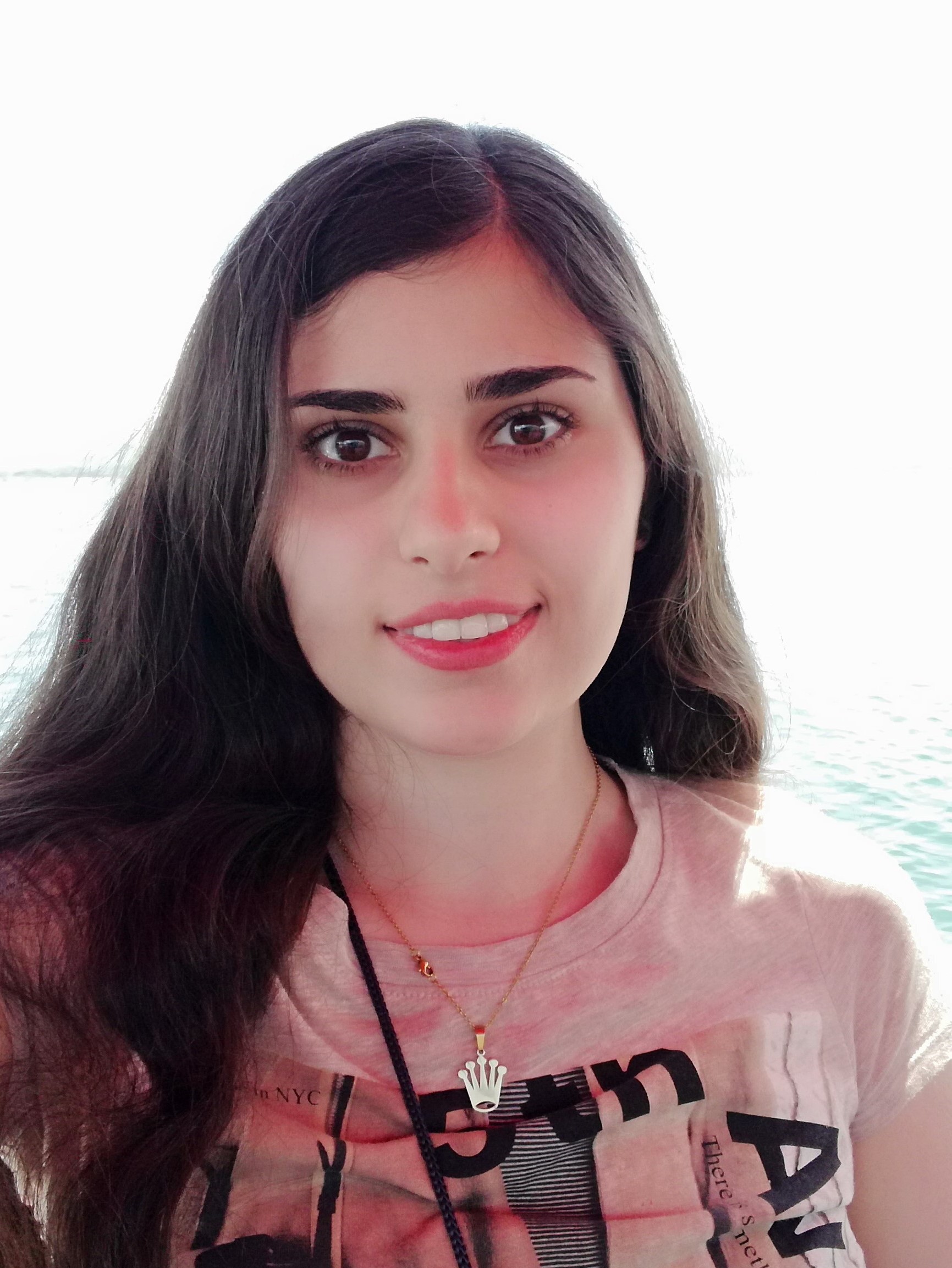}}]{Rozhin Yasaei}
Rozhin Yasaei received her B.Sc. degree in Electrical Engineering from Sharif University of Technology, Iran, in 2018. She received her M.Sc. degree in Computer Engineering from the University of California Irvine, USA, where she is currently pursuing her Ph.D. in Computer Engineering. Her research interests lie in the intersection of hardware and software. It revolves around the data-driven modeling of cyber-physical systems for automation and security using machine learning.
\end{IEEEbiography}
\vspace{-1cm}
\begin{IEEEbiography}[{\includegraphics[width=1in,height=1.25in,clip,keepaspectratio]{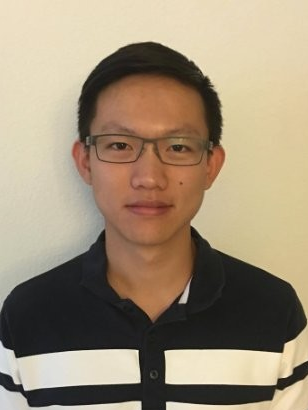}}]{Luke Chen}
Luke Chen received his B.S. (2019) in Electrical and Computer Engineering from University of California Irvine (UCI) in Irvine, CA, USA, where he is currently pursuing a Ph.D. degree in Electrical and Computer Engineering with a focus in Computer Engineering at UCI. His current research includes low-power embedded systems and applied machine learning in the areas of mobile health, autonomous systems, edge-cloud and split computing.
\end{IEEEbiography}
\vspace{-1cm}
\begin{IEEEbiography}[    {\includegraphics[width=1in,height=1.25in,clip,keepaspectratio]{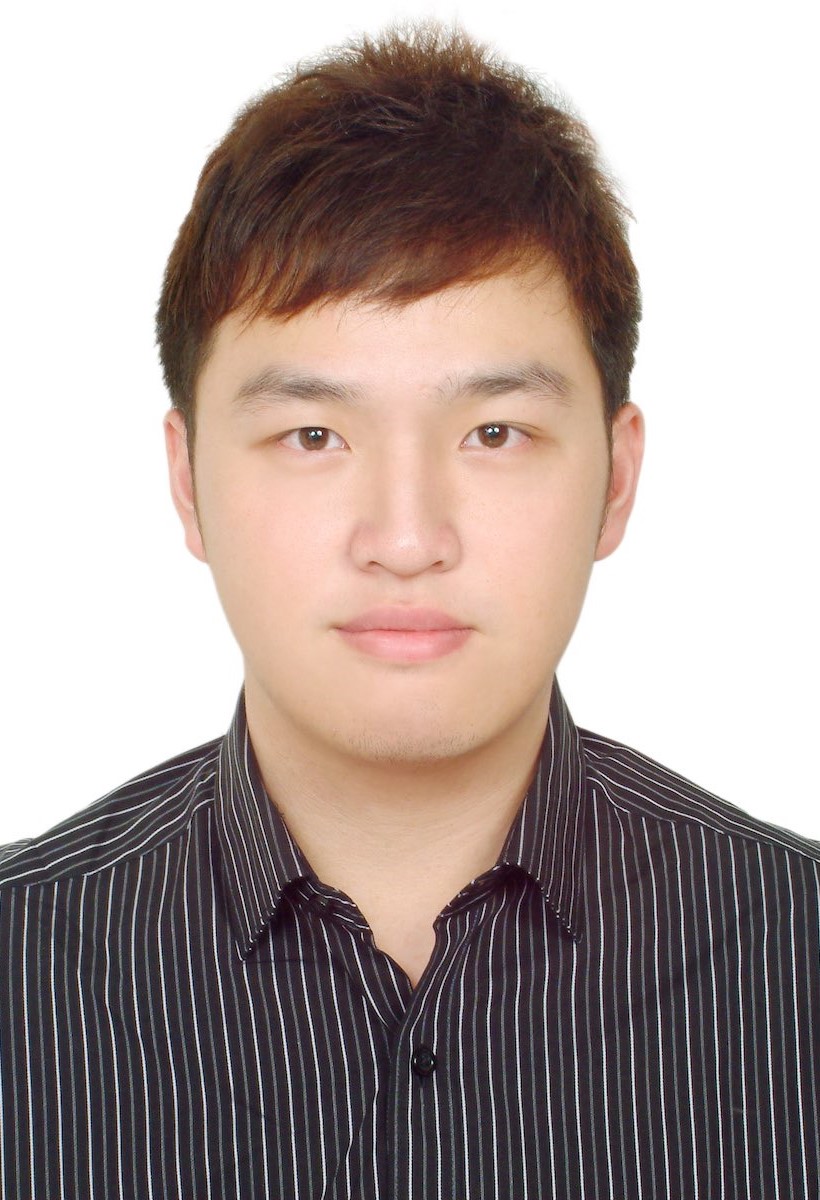}}]
{Shih-Yuan Yu} received the B.S. and M.S. degrees in Computer Science and Information Engineering from the National Taiwan University (NTU) in 2014. He worked at MediaTek for 4 years. Currently he is a Ph.D. student in the University of California, Irvine, USA. Now his research interests are about design automation of embedded systems using data-driven system modeling approaches.It covers incorporating machine learning methods to identify potential security issues in systems.
\end{IEEEbiography}
\vspace{-1cm}
\begin{IEEEbiography}[{\includegraphics[width=1in,height=1.25in,clip,keepaspectratio]{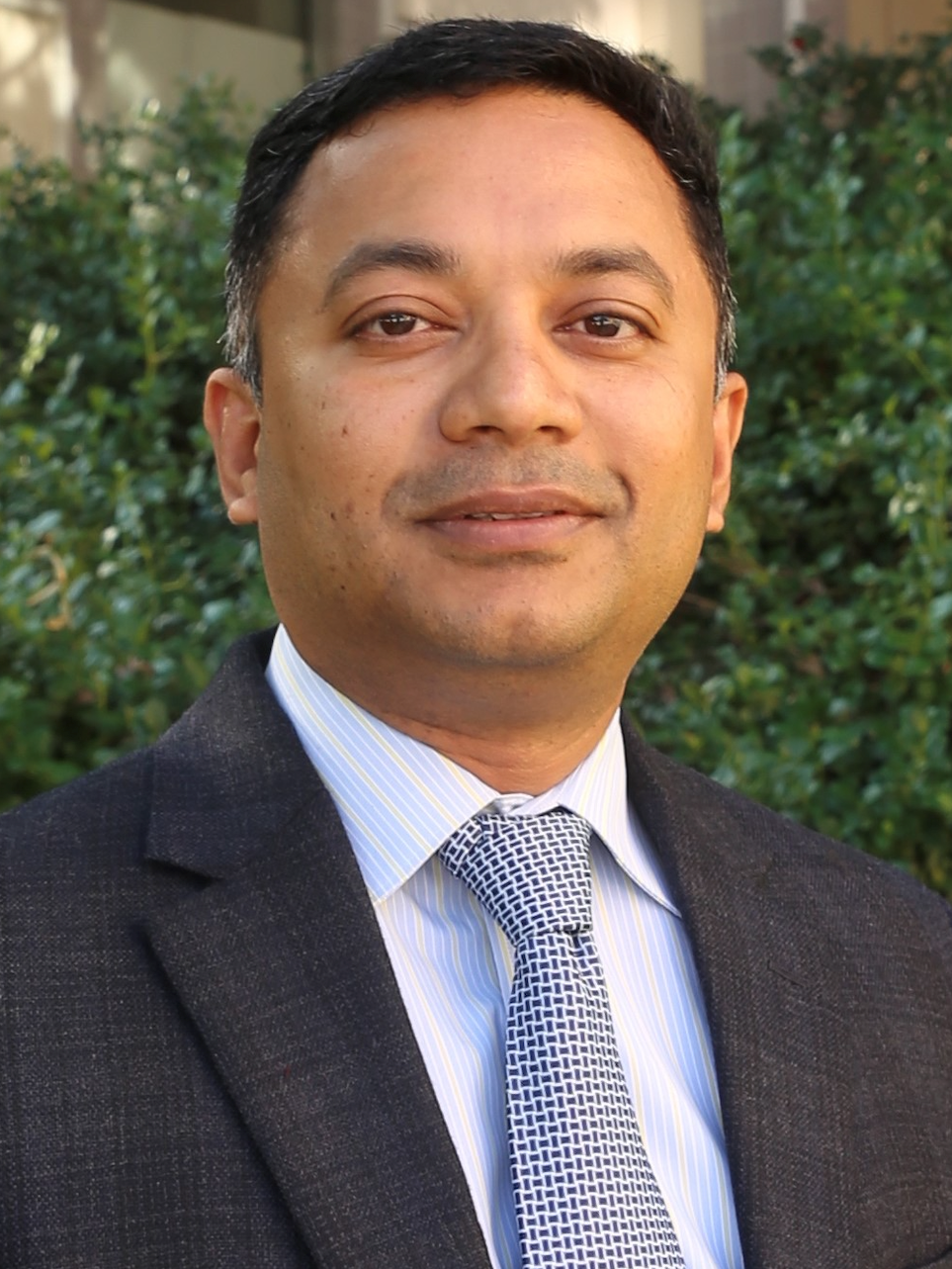}}
]{Mohammad Abdullah Al Faruque}
Mohammad Abdullah Al Faruque received his B.Sc. degree in Computer Science and Engineering (CSE) from Bangladesh University of Engineering and Technology (BUET) in 2002, and M.Sc. and Ph.D. degrees in Computer Science from Aachen Technical University and Karlsruhe Institute of Technology, Germany in 2004 and 2009, respectively. Mohammad Al Faruque is currently with the University of California Irvine (UCI), an associate professor and directing the Embedded and Cyber-Physical Systems Lab. Before he was with Siemens Corporate Research and Technology in Princeton, NJ. Prof. Besides 120+ IEEE/ACM publications in the premier journals and conferences, Prof. Al Faruque holds 11 US patents. Prof. Al Faruque is currently serving as the associate editors of the ACM Transactions on Design Automation on Electronics and Systems and the IEEE Design and Test. He is an IEEE senior member and an ACM senior member. He is also the IEEE CEDA Distinguished Lecturer for 2022-2023.

\end{IEEEbiography}
\vfill

\end{document}